\documentstyle[twoside,ijmpd,overcite]{article}  

\artsectnumbering

\def\Ref{Ref.~} 

\begin{document}

\runninghead{D.\ Bini, P.\ Carini and R.T.\ Jantzen}{Intrinsic Derivatives and Centrifugal Forces in General Relativity $\ldots$}

\normalsize\textlineskip
\thispagestyle{empty}
\setcounter{page}{1}

\copyrightheading{Vol.\ 6, No.\ 1 (1997) 1--38 [reformatted 2001]}

\vspace*{0.88truein}

\fpage{1}

  \centerline{\bf
THE INTRINSIC DERIVATIVE AND CENTRIFUGAL FORCES
  }\vspace*{0.035truein}
  \centerline{\bf
IN GENERAL RELATIVITY: I. THEORETICAL FOUNDATIONS
  }

  \vspace*{0.37truein}
  \centerline{
DONATO BINI
  }
  \vspace*{0.015truein}
  \centerline{\footnotesize\it 
Istituto per Applicazioni della Matematica C.N.R., 
I--80131 Napoli, Italy 
  and
  }
  \baselineskip=10pt
  \centerline{\footnotesize\it 
International Center for Relativistic Astrophysics, 
University of Rome, I--00185 Roma, Italy
  }
  \vspace*{10pt}
  \centerline{
PAOLO CARINI\footnote{
  Present Address: Physics Department, Amherst College, 
  Amherst, MA 01002, USA.}
  }
  \vspace*{0.015truein}
  \centerline{\footnotesize\it 
GP-B, Hansen Labs, Stanford University, 
Stanford, CA 94305, USA 
and
  }
  \baselineskip=10pt
  \centerline{\footnotesize\it 
International Center for Relativistic Astrophysics,
University of Rome, I--00185 Roma, Italy
  }
  \vspace*{10pt}
  \centerline{
ROBERT T. JANTZEN
  }
  \vspace*{0.015truein}
  \centerline{\footnotesize\it 
Department of Mathematical Sciences, 
Villanova University, Villanova, PA 19085, USA 
and
  }
  \baselineskip=10pt
  \centerline{\footnotesize\it 
International Center for Relativistic Astrophysics, 
University of Rome, I--00185 Roma, Italy
  }

  \vspace*{0.225truein}
  \pub{17 December 1996}

  \vspace*{0.21truein}


\abstracts{
Everyday experience with centrifugal forces has always guided thinking
on the close relationship between gravitational forces and accelerated
systems of reference. Once spatial gravitational forces and
accelerations are introduced into general relativity through a
splitting of spacetime into space-plus-time associated with a family
of test observers, one may further split the local rest space of those
observers with respect to the direction of relative motion of a test
particle world line in order to define longitudinal and transverse
accelerations as well. The intrinsic covariant derivative (induced
connection) along such a world line is the appropriate mathematical
tool to analyze this problem, and by modifying this operator to
correspond to the observer measurements, one understands more clearly
the work of Abramowicz et al who define an ``optical centrifugal
force'' in static axisymmetric spacetimes and attempt to generalize it
and other inertial forces to arbitrary spacetimes. In a companion
article the application of this framework to some familiar stationary
axisymmetric spacetimes helps give a more intuitive picture of their
rotational features including spin precession effects, and puts
related work of de Felice and others on circular orbits in black hole
spacetimes into a more general context. 
}{}{}

\vspace*{10pt}
\keywords{gravitoelectromagnetism--inertial forces}

\textlineskip
\setcounter{footnote}{0}
\renewcommand{\thefootnote}{\alph{footnote}}

\section{Introduction}

Ever since Einstein unified space and time into spacetime, people have
been trying to break them apart again. Spacetime splittings play an
important role in many aspects of gravitational theory, not only in
helping interpret 4-dimensional geometry in terms of our more familiar
space-plus-time perspective, but also in mathematical analysis of
various problems of the theory. Although there are many variations on
the idea of reintroducing space and time, all share a common
foundation of introducing a family of test observers in spacetime who
measure spacetime quantities mathematically by the orthogonal
decomposition of the tangent spaces to the spacetime manifold into
their local rest spaces and local time directions \cite{mfg,mit}. Such
a construction leads to a ``reference frame" or ``reference system" or
other variations of this terminology whose effect in general is to
contribute inertial forces to the spatial force equation due to the
motion of the family of test observers. Although one can agree on
these concepts in simple nonrelativistic situations, the richness of
general relativity and of the geometry of spacetime allows many
variations of their possible generalizations to the latter theory.
There is not necessarily a single ``correct" generalization of any
given concept, but simply different ways of measuring different
quantities, some of which may be more useful than others. 

Because of the equivalence principle, inertial forces have always
entrigued people in connection with gravitational theory. Indeed the
concept of centrifugal force is useful in nonrelativistic mechanics,
and in recent years Abramowicz et al
\cite{acl88,a90a,a90b,a90c,ab91,a92,ams93} have shown that a certain
generalization of this concept can give a nice physical interpretation
to certain properties of strong static gravitational fields in general
relativity, although attempts to extend it first to stationary and
then to arbitrary spacetimes have been somewhat problematic
\cite{pc90,ip93,a93,anw93,anw95,sem,nayvis96,pra96}. 
Here we place that work and related studies of de Felice and others
\cite{def91,defuss91,defuss93,def94,def95,barboiisr,sem96} in the more general
context of gravitoelectromagnetism, the framework which encompasses all the
various splitting approaches to general relativity and provides a clean
description of the possible choices of curved spacetime generalizations of
centripetal acceleration and centrifugal and Coriolis forces. This is done not
to exaggerate the importance of splitting spacetime but to help clarify the
link between our three-dimensional world view and nonrelativistic common
experience on the one hand and the interpretation of concepts related to
rotation and acceleration in general relativity on the other. 

\section{The nonrelativistic background}

Before exploring the spacetime picture, it is worth recalling the
classic example from nonrelativistic mechanics of a rigidly rotating
Cartesian coordinate system and its relation to the ideas of
centrifugal force and centripetal acceleration. It is here that all
our intuition for these concepts has its roots. 

Let $x^i= R^i{}_j(t) {\bar x}{}^j$ be the coordinate transformation
between ``space-fixed" orthonormal coordinates $\{\bar x{}^i \}$ and
rotating ``body-fixed" such coordinates $\{ x^i \}$ with a common
origin in Euclidean space, borrowing from the usual terminology of the
rigid body problem in classical mechanics. The body-fixed components
of the angular velocity of the rotating system are defined by 
\begin{equation}
  \dot R{}^i{}_k R^{-1}{}^k{}_j 
      = \delta^{ik}\epsilon_{kjm}\Omega^m \ .
\end{equation}
Both these and the space-fixed components $\bar{\Omega}{}^i =
R^{-1}{}^i{}_j \Omega^j$ are constant in the case of a rotation about
a fixed axis with constant angular velocity about that axis. 

If one evaluates the first and second time derivatives of the
coordinates of a trajectory, one easily finds 
\begin{eqnarray}
  \dot{\bar x}{}^i  
       &=& R^{-1}{}^i{}_j [ V^j + \dot{x}{}^j ] \ ,
                          \nonumber\\
  \ddot{\bar x}{}^i 
       &=& R^{-1}{}^i{}_j [ A^j + 2 (\Omega \times \dot{x} )^j                        + \ddot{x}{}^j ] \ ,
\end{eqnarray}
where $V^i = (\Omega \times x )^i = \delta^{ij}\epsilon_{jkm} \Omega^k
x^m$ is the relative velocity field of the body-fixed points relative
to the space-fixed points, and $A^i = (\Omega \times (\Omega \times x
))^i + (\dot\Omega \times x)^i$ is the relative acceleration field,
both evaluated along the trajectory. 

If $\ddot{\bar x}{}^i = {\bar F}{}^i$ is the space-fixed force per
unit mass equation of motion for a point particle of mass $m$ moving
under the influence of a force $m{\bar F}{}^i$, then the acceleration
relation above may be rewritten in the form 
\begin{equation}
   \ddot{x}{}^i = F^i + g^i + (\dot{x} \times H)^i \ ,
\end{equation}
by moving the additional terms due to the time dependence of the
transformation to the other side of the equation where they appear as
\lq\lq inertial forces" due to the motion of the body-fixed points to
which the coordinate system is attached. One has a
``gravitoelectric-like" force $g^i = - A^i$ due to the acceleration of
those points, and the remaining \lq\lq gravitomagnetic-like" force,
the Coriolis force, which is due to the changing orientation of the
body-fixed axes, which in turn is a manifestation of the relative
motion of the body-fixed points. The latter force involves a
``gravitomagnetic-like field" $H^i = 2 \Omega^i$. The
``gravitoelectric-like'' force consists of the centrifugal force
directed away from the axis of rotation and an additional force 
due to the changing angular velocity.

In the more familiar case of a time-independent angular velocity,
these two noninertial force fields admit a scalar and a vector
potential respectively. The scalar potential is just half the square
of the magnitude of the relative velocity field 
\begin{eqnarray}
   \Phi &=& {\textstyle 1\over 2} \delta_{ij} V^i V^j \ ,
            \nonumber\\
   g^i  &=& - \delta^{ij} \mathop{\rm grad}\nolimits_j \Phi \ ,
\end{eqnarray}
while this field itself serves as the vector potential
\begin{equation}
    H^i = (\mathop{\rm curl} V)^i \ .
\end{equation}
Since each body-fixed point is undergoing circular motion at constant
velocity, the relative acceleration field is exactly the familiar
centripetal acceleration associated with this simple motion, so the
centrifugal force is just the sign-reversal of this centripetal
acceleration. The Coriolis force can be interpreted as due to the
local vorticity of the flow of the body-fixed points in space (half
the curl of the velocity field), which is equal to the global constant
angular velocity vector $\Omega^i$. 

\section{The Spacetime Setting}

The great simplication of this discussion compared to a corresponding
one in terms of a geometric splitting of spacetime is the common
Newtonian time used by both systems of spatial coordinates. In a
spacetime discussion, one must also take into account the change in
the local time direction, which complicates matters, especially the
relationship between a sign-reversed centripetal acceleration and a
corresponding centrifugal force. One also must re-interpret the time
derivative in a way which makes geometric sense, and there are a
number of distinct ways of doing this, depending on how changes in
fields are measured along general world lines. 

Consider only the case of time-independent angular velocity. By adding
the additional time coordinate transformation 
\begin{equation}
    t = \bar t
\end{equation}
to the original spatial coordinate transformation, one obtains a
transformation from inertial coordinates $\{\bar t,\bar x{}^i\}$ in
Minkowski spacetime to noninertial coordinates $\{t,x^i\}$ which may
be interpreted using the spacetime geometry. The time lines of each
such coordinate system sharing the same time coordinate hypersurfaces
in Minkowski spacetime may be interpreted as the world lines of a
family of test observers (when timelike), representing the
trajectories of the space-fixed and body-fixed points. The first set
are inertial (zero acceleration) observers with zero relative
velocity, whose world lines are the orthogonal trajectories to the
family of time hypersurfaces, while the second are noninertial
(accelerated) observers in relative motion and not admitting any
orthogonal family of hypersurfaces. Both families of world lines are
the flow lines of Killing vector fields of Minkowski spacetime
(therefore having zero expansion tensor), the first with zero
vorticity and the second with nonzero vorticity. 

In this description, it is only the families of time lines and time
hypersurfaces which play a central role, not the specific choice of
spatial coordinates which parametrize the families of time lines. It
is in fact useful to introduce new spatial coordinates adapted to the
orbits of the body-fixed points, and their corresponding rotating
system, namely nonrotating and rotating cylindrical coordinates
adapted to the axis of the rotation. For example, if one chooses the
axis of rotation so that $\bar{\Omega}^i = \Omega^i = \Omega
\delta^i{}_3$, one can choose the usual cylindrical coordinates
$\{\bar{\rho},\bar{\phi},\bar{z}\}$ in place of $\{{\bar x}{}^i\}$,
and rotating cylindrical coordinates $\{\rho,\phi,z\}$ which differ
only by $\phi = \bar{\phi} - \Omega \bar t$. 

However, a description in terms of quantities measured by each family
of test  observers involves their local proper times, which in the
second case are not associated with any global time function. Global
time functions not directly measuring observer proper time occur in
the context of ``observer-adapted" coordinate systems. A system of
coordinates can be adapted to the observer congruence in one of two
ways \cite{mfg}. For a general congruence with nonzero vorticity, the
appropriate coordinates are comoving, leading to an approach called
the threading point of view, while for a vorticity-free congruence,
one can use more general coordinates adapted to the family of
orthogonal hypersurfaces admitted by the congruence, leading to an
approach called the slicing point of view. In each case, the adapted
local coordinates $\{t,x^i\}$ lead to the introduction of explicit
potentials for the various gravitational forces. 

A slicing together with a transversal threading (congruence) describes
exactly the structure on which each of these two points of view is
built, here called a ``nonlinear reference frame." In the flat
Minkowski space example with inertial or rotating observers, one has
the geodesically parallel slicing orthogonal to the paths of the
original inertial observers, together with the rotating observers
which provide a new threading of this slicing (when those observers
are defined). Rotating Cartesian or cylindrical coordinates on
Minkowski spacetime are examples of coordinates adapted to the
nonlinear reference frame associated with this slicing and threading.
In the threading point of view the threading congruence serves both as
the observer congruence as well as the curves along which evolution is
measured, while in the slicing point of view these two roles are
separated: the orthogonal trajectories to the slicing form the
observer congruence, while the measured quantities are evolved along
the distinct threading congruence. 

This is a nonlinear reference frame adapted to the stationary
axisymmetry of the flat Minkowski spacetime. A similar geometrically
privileged nonlinear reference frame exists in any stationary
axisymmetric spacetime, in particular in the Kerr black hole
spacetimes (with the Schwarzschild black hole as their static limit)
and in the G\"odel spacetime. For such spacetimes with some of the
same symmetries which characterize the uniformly rotating observers in
Minkowski spacetime, one might hope to define generalized centrifugal
and Coriolis forces, but without these symmetries one must rethink the
point of departure. 

This is not the whole story about centrifugal force in classical
mechanics, since it makes its appearance in at least two other
familiar contexts. As discussed in detail by Abramowicz \cite{a90b},
one is the train, plane, car context in which one has an accelerated
platform to which a local reference frame is attached, essentially the
previous problem with additional motion of the origin of coordinates.
Any point fixed in this local platform will then experience
accelerations tangential and transverse to its direction of motion,
and the transverse acceleration can be interpreted in terms of a
centrifugal force in the local reference frame due to its
instantaneous rotation about the center of the osculating circle
associated with the curvature of its path. If the point is in motion
with respect to the local platform, Coriolis effects are felt as well,
but the details are more complicated than the simpler rigid body
discussion. 

A third context in which the centrifugal force is usually introduced
is in the discussion of motion in a central potential \cite{a90b}.
Here one introduces a polar coordinate system in the plane of the
motion and then expresses the equation of motion in that coordinate
system. The radial component of this equation for general motion then
contains what is interpreted as a centrifugal force term due to the
curvature of the circular angular coordinate lines. This term is just
the sign-reversal of the centripetal acceleration for motion confined
to these coordinate lines and enters the equation of motion through a
Christoffel symbol term associated with this curvature, quadratic in
the angular speed. Although there is no rotating frame in this
discussion, one may be introduced by letting the new system rotate
about the center of force so that the particle in motion has a fixed
new angular coordinate. In this rotating frame, the same centrifugal
force term is then realized as in the rigid body discussion as a
rotating frame effect, but rotating about the center of force, not the
instantaneous radial direction associated with the curvature of the
particle path. For circular motion, this centrifugal force is just the
sign-reversal of the centripetal acceleration of the particle path,
but for general noncircular motion, the two quantities are not simply
linked. 

In other words, the ``fictitious'' centrifugal force is a convenience
that only has meaning with respect to some implied reference frame,
and in the same problem can play different roles depending on which
frame is chosen. For circular trajectories, all of these various
aspects of centrifugal force come together, presenting the most useful
application of the concept. 

In general relativity the best hope of having a useful generalization
of centrifugal force lies in the static axisymmetric case with
circular trajectories. Consideration of noncircular trajectories in
that case or relaxing the symmetry even to stationarity already
introduces difficulties which make the whole discussion rather
unclear. However, the idea of a relative centripetal acceleration
viewed by an observer, being well defined in a single reference frame,
sidesteps the questions involving two different reference frames that
seem to be tied up with various aspects of centrifugal force and it is
therefore reasonable to generalize it to an arbitrary spacetime. 

The key difference between the nonrelativistic description of inertial
forces and the framework of general relativity is that in the latter
context, the effects of a gravitational field due to the presence of 
matter are intertwined with those of the accelerated motion of the
field of observers used to establish a given reference frame in order
to ``measure" the gravitational forces. Only under special symmetry
conditions does it seem to make sense to try to separate the two.
Moreover, focusing attention on a single test particle world line
without referring it to an independent family of test observers (which
are not adapted to the particular worldline) reduces the usefulness of
a space-plus-time interpretation of the motion, except possibly in
reference to the spacetime Frenet-Serret frame which is completely
determined by the world line alone
\cite{honschvis,greschvis,iyevis88,iyevis93}. 

\section{A family of test observers}

A splitting of a general spacetime equipped with a Lorentzian metric
$g_{\alpha\beta}$ (signature {\tt -+++}) and the covariant derivative
$\nabla_\alpha$ associated with its symmetric connection is
accomplished locally by specifying a future-pointing unit timelike
vector field $u^\alpha$ ($u^\alpha u_\alpha = -1$). This field may be
interpreted as the four-velocity of a family of test observers whose
proper time parametrized world lines (let $\tau_u$ denote such a
parameter on each world line) are integral curves of $u^\alpha$. 

The orthogonal decomposition of each tangent space into the local rest
space and local time direction of the observer extends to all the
tensor spaces above it and to the algebra of spacetime tensor fields,
and may be referred to as the measurement process associated with the
family of test observers. Tensors or tensor fields which have no
component along $u^\alpha$ are called spatial (with respect to
$u^\alpha$). The fully covariant and contravariant forms of the
spatial projection tensor $P(u)^\alpha{}_\beta = \delta^\alpha{}_\beta
+ u^\alpha u_\beta$ are referred to as the corresponding forms of the
spatial metric 
\begin{equation}\label{proj}
     P(u)_{\alpha\beta} = g_{\alpha\beta} + u_\alpha u_\beta \ .
\end{equation}
Similarly let $\eta(u)_{\alpha\beta\delta} =
u^\delta\eta_{\delta\alpha\beta\delta}$ be the spatial unit
antisymmetric tensor associated with the spatial metric, where
$\eta_{0123} = 1 = \eta(u)_{123}$ in a time-oriented, oriented
orthonormal frame having $u^\alpha$ as its first element. This tensor
may be used to define a spatial duality operation for antisymmetric
spatial tensor fields in an obvious way. 

As described in \Ref\citen{mfg}, one may also spatially project various
derivative operators so that the result of the derivative of any
tensor field is always spatial; such derivatives naturally occur in
expressing tensor equations in space-plus-time form. Two useful
spatial derivatives are the spatial Lie derivative $
\pounds(u)_{\hbox{$X$}} = P(u) \pounds_{\hbox{$X$}}$, for any spatial
vector field $X^\alpha$, and the spatial covariant derivative $
\nabla(u)_\alpha = P(u) P(u)^\beta{}_\alpha \nabla_\beta$, where the
projection on all free indices is implied after the application of the
derivative. Similarly three useful temporal derivatives are the
temporal Lie derivative $ \nabla_{\rm(lie)}(u) = P(u)
\pounds_{\hbox{$u$}}$ , the Fermi-Walker temporal derivative $
\nabla_{(\rm fw)}(u) = P(u) u^\alpha\nabla_\alpha$, and the corotating
Fermi-Walker temporal derivative $\nabla_{\rm(cfw)}(u)$ related to the
first two by the kinematical fields (acceleration $a(u)^\alpha$,
vorticity $\omega(u)^\alpha{}_\beta$, expansion
$\theta(u)^\alpha{}_\beta$) of the observer congruence, 
\begin{eqnarray}
  a(u)^\alpha &=&  \nabla_{(\rm fw)}(u) u^\alpha \ ,
     \nonumber\\
 \omega(u)_{\alpha\beta} &=& P(u)^\gamma{}_\alpha P(u)^\delta{}_\beta
       \nabla(u)_{[\gamma} u_{\delta]} \ ,
     \nonumber\\
 \theta(u)_{\alpha\beta} &=& P(u)^\gamma{}_\alpha P(u)^\delta{}_\beta
       \nabla(u)_{(\gamma} u_{\delta)}
      \nonumber\\
     &=& \frac12 \nabla_{\rm (lie)}(u) g_{\alpha\beta}
        = \frac12 \nabla_{\rm (lie)}(u) P(u)_{\alpha\beta} \ ,
\end{eqnarray}
through the relations
\begin{equation} 
\nabla_{\rm(cfw)}(u) X^\alpha 
       = \nabla_{\rm(fw)}(u) X^\alpha 
           + \omega(u)^\alpha{}_\beta X^\beta 
       = \nabla_{\rm(lie)}(u) X^\alpha 
           + \theta(u)^\alpha{}_\beta X^\beta
\end{equation}
valid only for a spatial vector field $X^\alpha$, but easily extended
to any spatial tensor field in the usual way. It is also convenient to
use extend the notation $\pounds(u)_{\hbox{$X$}} =
P(u)\pounds_{\hbox{$X$}}$ to any vector field $X^\alpha$, spatial or
not. All indices are spatially projected when these spatial
differential operators are applied to a tensor field. The geometrical
meaning of these derivatives is discussed in \Ref\citen{mfg}. For spatial
fields, the Fermi-Walker temporal derivative coincides with the
spacetime Fermi-Walker derivative along the observer congruence,
explaining the terminology. 

Note that the spatial covariant derivative and the ordinary and
corotating Fermi-Walker temporal derivatives of the spatial metric are
all zero, so index-shifting of spatial fields commutes with these
derivatives. The Lie temporal derivative of the spatial metric instead
equals twice the expansion tensor of the observer congruence, so index
shifting of a spatial tensor being differentiated by this operator
leads to additional terms involving this tensor. 

It is useful to introduce a vorticity or rotation vector field using
the spatial duality operation 
\begin{equation}
     \omega(u)^\alpha = 
       {\textstyle {1\over2}} 
         \eta(u)^{\alpha\beta\gamma} \omega(u)_{\beta\gamma} \ .
\end{equation}
This in turn may be used to rewrite the contraction of the vorticity
tensor with a spatial vector field as a spatial cross-product 
\begin{equation}
     \omega(u)^\alpha{}_\beta X^\beta 
       = -\eta(u)^\alpha{}_{\beta\gamma} \omega(u)^\beta X^\gamma
       = - [ \vec\omega(u) \times_u X ]^\alpha \ .
\end{equation}
Finally the shear tensor is defined as the spatial tracefree part of the
expansion tensor
\begin{equation}
  \sigma(u)_{\alpha\beta} 
   = \theta(u)_{\alpha\beta} 
        - {\textstyle\frac13} \theta(u)^\gamma{}_\gamma
          P(u)_{\alpha\beta} \ .
\end{equation}

\section{Measuring the intrinsic derivative along a parametrized
         curve} 

Given any parametrized curve in spacetime, with parameter $\lambda$
and tangent $V(\lambda)^\alpha$, the spacetime connection induces a
connection on the curve whose derivative is called either the
``intrinsic" or ``absolute" derivative along it \cite{sacwu}. This
derivative $D/d\lambda$ is uniquely defined by the condition that if
one extends a tensor smoothly off the curve to a tensor field on
spacetime, the action of this intrinsic derivative on the tensor at a
point on the curve equals the action of the covariant directional
derivative $V(\lambda)^\alpha \nabla_\alpha$ on the extended tensor
field at that point 
\begin{equation}
      D T^{\alpha\ldots}_{\ \ \beta\ldots}/ d \lambda
       = V(\lambda)^\gamma  \nabla_\gamma 
                  T^{\alpha\ldots}_{\ \ \beta\ldots} \ .
\end{equation}
In this equation it is important to note that its right hand side is
understood to be the value on the worldline of the derivative of the
extended tensor field. The usual sloppy notation of this equation
which does not distinguish between the original and the extended
tensor field nor indicate its validity only on the curve itself must
always be understood in this context. 

For a vector defined only along the curve, this leads to the usual
formula in terms of the ordinary parameter derivative of the
components along the world line plus the connection coefficient
correctional terms 
\begin{equation}
         D X^\alpha / d \lambda
           = d X^\alpha / d \lambda 
      + \Gamma^\alpha{}_{\beta\gamma} V(\lambda)^\beta X^\gamma \ .
\end{equation}
In a previous article \cite{mfg} this derivative has been referred to
as the total covariant derivative along $V(\lambda)^\alpha$ or along
the parametrized curve. 

If one performs an orthogonal projection of the intrinsic derivative
along such a parametrized curve in the measurement process associated
with the family of test observers, one is led to a new derivative
operator along that curve which is natural to call either the
Fermi-Walker spatial intrinsic derivative or the Fermi-Walker total
spatial covariant derivative. For example, if $X^\alpha$ is any vector
defined along the curve, this derivative is defined by 
\begin{equation}\label{eq:Dfwdef}
      D_{\rm(fw)}(V(\lambda),u)X^\alpha / d \lambda 
        = P(u)^\alpha{}_\beta D X^\beta / d \lambda \ .
\end{equation}
If one extends $X^\alpha$ to a smooth vector field off the curve, then
for this extended vector field, expressing this derivative in terms of
the measurement of the tangent vector itself 
\begin{equation}
     V(\lambda)^\alpha 
         = V(\lambda)^{(||_u)} \, u^\alpha + [P(u) V(\lambda)]^\alpha
\end{equation}
leads to the natural pairing of its temporal part $V(\lambda)^{(||_u)}
= -u_\gamma V(\lambda)^\gamma$ and spatial part
$[P(u)V(\lambda)]^\alpha = P(u)^\alpha{}_\gamma V(\lambda)^\gamma$
with the corresponding projections of the covariant derivative acting
on the smooth extension 
\begin{equation}
      D_{\rm(fw)}(V(\lambda),u)X^\alpha / d \lambda
        = \{ V(\lambda)^{(||_u)} \nabla_{\rm(fw)}(u) 
          + [P(u) V(\lambda)]^\beta \nabla(u)_\beta \} X^\alpha \ ,
\end{equation}
where the individual terms on the right hand side depend on how the
extension is made. These individual terms have no meaning without such
an extension, a fact which has caused some confusion in attempts to
generalize centrifugal and other noninertial forces to curved
spacetimes \cite{a93,anw93,anw95,sem}. 

By replacing the Fermi-Walker temporal derivative in the expression
for the Fermi-Walker intrinsic derivative acting on the extended
tensor field with the Lie temporal derivative or the corotating
Fermi-Walker temporal derivative respectively, one defines the
corresponding total spatial covariant derivatives of a spatial vector
field 
\begin{eqnarray}
      D_{\rm(tem)}(V(\lambda),u)X^\alpha / d \lambda
        &=& \{ V(\lambda)^{(||_u)} \nabla_{\rm(tem)}(u) 
          + [P(u) V(\lambda)]^\beta \nabla(u)_\beta \} X^\alpha \ 
            \nonumber\\ 
        & &\quad {\scriptstyle\rm tem = fw,\, cfw,\, lie} \ ,
\end{eqnarray}
where again the righthand side expression only makes sense for an
extended vector field, but leads to well-defined derivatives for a
spatial vector defined only on the parametrized curve. 

These three derivatives of spatial vectors along the world line differ
among themselves only by a linear transformation of the local rest
space 
\begin{eqnarray}
      D_{\rm(cfw)}(V(\lambda),u)X^\alpha / d \lambda
      &=& D_{\rm(fw)}(V(\lambda),u)X^\alpha / d \lambda 
           +  V(\lambda)^{(||_u)} \omega(u)^\alpha{}_\beta X^\alpha
       \nonumber\\
      &=& D_{\rm(lie)}(V(\lambda),u)X^\alpha / d \lambda 
           +  V(\lambda)^{(||_u)} \theta(u)^\alpha{}_\beta X^\alpha
     \ ,
\end{eqnarray}
expressions which may be used together with Eq.~(\ref{eq:Dfwdef}) to
define the corotating and Lie such derivatives in terms of the
Fermi-Walker one when acting on spatial fields. The ordinary and
corotating Fermi-Walker total spatial covariant derivatives of the
spatial metric vanish so they commute with index shifting on spatial
fields, and one may introduce spatial orthonormal triads along the
curve for which one of these derivatives vanishes. These are natural
to call ``relative Fermi-Walker" or ``relative corotating
Fermi-Walker" propagated spatial frames along the parametrized curve.
The Lie such derivative of the spatial metric vanishes only if the
expansion tensor vanishes, in which case it coincides with the
corotating Fermi-Walker total spatial covariant derivative. Thus in
general the ``relative Lie" propagated spatial frames along the
parametrized curve will not remain orthonormal if initially so when the
expansion tensor is nonzero. 

The three different choices of derivative correspond to the three
possible ways of evolving spatial frames into the future, the first
two of which preserve inner products. They each measure differences
with respect to the associated relative propagated spatial frames
along the given curve. The relationship between the relative Lie
transport and the relative Fermi-Walker transport along the observer
congruence itself leads to the physical intepretation of the rotation,
expansion, and shear of that congruence. 

\section{Reparametrization of a parametrized curve}

Two new parametrizations may be introduced for any parametrized curve
by solving the following ordinary differential equations 
\begin{equation}
     d\tau_{(V(\lambda),u)} / d \lambda = V(\lambda)^{(||_u)} \ , \quad
     d\ell_{(V(\lambda),u)} / d \lambda = ||P(u)V(\lambda)|| \ ,
\end{equation}
(using the notation $||Y|| = |Y_\alpha Y^\alpha|^{1/2}$) corresponding
to the limiting sequence of temporal and spatial arclength
differentials seen by the test observers whose paths are crossed by
the curve. The solutions lead to valid reparametrizations as long as
the right  hand side of the differential equation does not vanish, so
that an invertible relationship exists between the old and new
parametrizations. When one of the two right hand sides vanishes
identically, the parameter becomes a proper interval parameter (proper
distance orthogonal to the observer family or proper time along it
respectively). 

The derivatives of these new parameters are in turn related to the
derivatives of the spacetime interval by the usual relation 
\begin{equation}
      [ d s / d \lambda ]^2 
         = - [ d \tau_{(V(\lambda),u)} / d \lambda ]^2
           + [ d \ell_{(V(\lambda),u)} / d \lambda ]^2 \ ,
\end{equation}
while their quotient up to sign defines the relative speed of the
curve as seen by the observer family 
\begin{equation}
   \pm \nu(V(\lambda),u)
      =  d\ell_{(V(\lambda),u)} / d\tau_{(V(\lambda),u)} 
      =  ||P(u)V(\lambda)|| / V(\lambda)^{(||_u)} \ .
\end{equation}
The relative velocity itself and (when nonzero) the unit vector
defining its direction are themselves defined by 
\begin{equation}
     \nu(V(\lambda),u)^\alpha 
         = [ P(u)V(\lambda) ]^\alpha / V(\lambda)^{(||_u)} \ ,
\end{equation}
and 
\begin{equation}
    \hat\nu(V(\lambda),u)^\alpha 
        = [ P(u)V(\lambda) ]^\alpha / ||P(u)V(\lambda)|| \ .
\end{equation}
The total spatial covariant derivatives along the parametrized curve
may be re-expressed in terms of the new parametrizations by the chain
rule 
\begin{equation}
    D/d\lambda' = [d \lambda / d\lambda'] \, D/ d\lambda 
                = [d \lambda' / d\lambda]^{-1} \, D/ d\lambda \ ,
\end{equation}
where $D$ here stands for any of the intrinsic derivative operators
and $\lambda'$ for either new parametrization. 

Since the Fermi-Walker and corotating Fermi-Walker such derivatives
respect spatial orthogonality, one may use them to introduce a
relative spatial Frenet-Serret frame of each type along the
parametrized curve, using the relative spatial arclength
parametrization to generalize the usual objects on a Riemannian
3-manifold. The spatial unit vector $\hat\nu(V(\lambda),u)^\alpha$
plays the role of the ``relative unit tangent," and the direction and
length of its derivative with respect to the relative spatial
arclength yields the ``relative unit normal" and ``relative spatial
curvature" of each type, while the spatial cross-product of the
relative unit tangent and normal defines the ``relative unit
bi-normal," whose derivative in turn leads to the ``relative torsion."
For a stationary spacetime with test observers following the
trajectories of timelike Killing vector field, the relative
Frenet-Serret structure for the corotating case corresponds to the
usual such structure on the observer quotient space with the projected
Riemannian metric, suggesting that it is a generalization worth
considering. Such a relative spatial Frenet-Serret structure should be
clearly distinguished from the spacetime Frenet-Serret structure for
the curve \cite{honschvis,greschvis,iyevis88,iyevis93}. 
 
\section{Measuring the intrinsic derivative along a test particle
         world line} 

Suppose one considers the world line of a nonzero rest mass test
particle in spacetime parametrized by the particle's proper time
$\tau_U$, letting $U^\alpha$ denote its unit timelike tangent vector,
the four-velocity of the test particle. This vector may be split using
the orthogonal decomposition associated with the family of test
observers with four-velocity $u^\alpha$ 
\begin{equation}
   U^\alpha = \gamma(U,u) [ u^\alpha + \nu(U,u)^\alpha ]
             = E(U,u)  u^\alpha + p(U,u)^\alpha  \ .
\end{equation}
Here the spatial vector $\nu(U,u)^\alpha$ is the relative velocity of
$U^\alpha$ with respect to $u^\alpha$, and $\gamma(U,u) = [ 1 -
\nu(U,u)^2 ]^{-1/2}$ is its associated gamma factor, while $\nu(U,u) =
[\nu(U,u)_\alpha \nu(U,u)^\alpha]^{1/2}$ is the 
relative speed. Similarly $p(U,u)^\alpha = \gamma(U,u)
\nu(U,u)^\alpha$ is the three-momentum (per unit mass) observed by the
test observers, with magnitude $p(U,u)$, while $E(U,u) = \gamma(U,u)$
is the energy (per unit mass) as seen by the test observers. (The
tilde notation of \Ref\citen{mfg} used for per unit mass quantities will be
dropped for simplicity.)  Either set of quantities satisfies the
identity 
\begin{equation}\label{eq:emc2}
    \gamma(U,u)^2 = \gamma(U,u)^2 \nu(U,u)^2 + 1\ , \quad
    E(U,u)^2 =  p(U,u)^2 + 1\ ,
\end{equation}
imposed by the unit nature of $u^\alpha$. Finally the two new relative
parametrizations of the world line are here defined by 
\begin{equation}
    d \tau_{(U,u)} / d \tau_U = \gamma(U,u) \ , \quad
    d \ell_{(U,u)} / d \tau_U = \gamma(U,u) \nu(U,u) \ ,
\end{equation}
with
\begin{equation}\label{eq:elltau}
    d \ell_{(U,u)} / d \tau_{(U,u)} = \nu(U,u) \ .
\end{equation}

Adopting the observer proper time parametrization, the spatial
projection of the total covariant derivative along the world line
$D/d\tau_{(U,u)} = \gamma(U,u)^{-1} D/d\tau_U$ defines the
Fermi-Walker total spatial covariant derivative, which together with
its two spatial generalizations can be expressed in the following way
for a spatial vector field $X^\alpha$ defined along the world line and
which has been extended off the world line for the right hand side to
make sense 
\begin{equation}
      D_{\rm(tem)}(U,u)X^\alpha / d \tau_{(U,u)} 
       = [\nabla_{\rm(tem)}(u) 
          + \nu(U,u)^\beta \nabla(u)_\beta ]X^\alpha \ ,
         \quad {\scriptstyle\rm tem = fw,\, cfw,\, lie} \ .
\end{equation}
These three derivatives of spatial vectors along the world line differ
among themselves only by a linear transformation of the local rest
space 
\begin{eqnarray}
      D_{\rm(cfw)}(U,u)X^\alpha / d \tau_{(U,u)} 
      &=& D_{\rm(fw)}(U,u)X^\alpha / d \tau_{(U,u)} 
                        + \omega(u)^\alpha{}_\beta X^\alpha
             \nonumber\\
      &=& D_{\rm(lie)}(U,u)X^\alpha / d \tau_{(U,u)} 
                        + \theta(u)^\alpha{}_\beta X^\alpha
     \ ,
\end{eqnarray}
expressions which may be used to define the Lie and corotating
Fermi-Walker such derivatives in terms of the Fermi-Walker one when
acting on spatial fields. 

\section{Relative acceleration: longitudinal and transverse parts}

Applying these derivatives to the relative velocity itself leads to a
relative acceleration vector for each one 
\begin{equation}
     a_{\rm(tem)}(U,u)^\alpha 
        = D_{\rm(tem)} \nu(U,u)^\alpha / d \tau_{(U,u)} \ ,
\end{equation}
differing from each other in the same way as the above three
derivatives 
\begin{equation}\label{eq:aaa}
         a_{\rm(cfw)}(U,u)^\alpha 
       = a_{\rm(fw)}(U,u)^\alpha  
               + [ \vec\omega(u) \times_u \nu(U,u) ]^\alpha
       = a_{\rm(lie)}(U,u)^\alpha  
               + \theta(u)^\alpha{}_\beta \nu(U,u)^\beta \ .
\end{equation}
The first of these equations just reflects the relative rotation of
the relative Fermi-Walker and corotating Fermi-Walker transported axes
along the world line. Apart from a gamma factor, the rate of change of
the spatial momentum is related to the relative acceleration  by an
additional term along the direction of relative motion involving the
rate of change of the energy (per unit mass) $E(U,u) = \gamma(U,u)$ of
the test particle 
\begin{equation}
     D_{\rm(tem)}(U,u) p(U,u)^\alpha / d\tau_{(U,u)} 
        = \nu(U,u)^\alpha d \ln \gamma(U,u) / d \tau_{(U,u)}
          +  \gamma(U,u)  a_{\rm(tem)}(U,u)^\alpha \ .
\end{equation}

One may further decompose both the relative acceleration and the
observed rate of change of spatial momentum into longitudinal and
transverse components with respect to the observed motion of the test
particle using the relative motion projectors \cite{mfg,rok} 
\begin{equation}
    P_u(U,u)^{(||)}{}^\alpha{}_\beta 
        = \hat\nu(U,u)^\alpha \hat\nu(U,u)_\beta\ ,
        \quad
    P_u(U,u)^{(\bot)}{}^\alpha{}_\beta 
        = P(u)^\alpha{}_\beta -  P_u(U,u)^{(||)}{}^\alpha{}_\beta \ .
\end{equation}
For the ordinary and corotating Fermi-Walker cases, this decomposition
is equivalent to the terms arising from the product rule when these
quantities are represented as the scalar product of their magnitude
and direction 
\begin{equation}
     \nu(U,u)^\alpha =\nu(U,u) \hat\nu(U,u)^\alpha \ ,\quad
     p(U,u)^\alpha = p(U,u) \hat\nu(U,u)^\alpha \ ,
\end{equation}
where $p(U,u) = \gamma(U,u) \nu(U,u)$. Consider first the relative
accelerations, which decompose into two terms 
\begin{eqnarray}
   a_{\rm(tem)}(U,u)^\alpha 
       &=& \hat\nu(U,u)^\alpha d \nu(U,u) / d \tau_{(U,u)}
   + \nu(U,u) D_{\rm(tem)}(U,u) \hat\nu(U,u)^\alpha / d \tau_{(U,u)}
      \nonumber\\
       &=&    a_{\rm(tem)}^{(||)}(U,u)^\alpha  
          +   a_{\rm(tem)}^{(\bot)}(U,u)^\alpha \ ,\quad
              {\rm\scriptstyle tem = fw,\, cfw} \ ,
\end{eqnarray}
which define respectively their components parallel (``tangential" to
the observed orbit, or longitudinal) and perpendicular (``normal" or
transverse) to the relative direction of motion, most naturally called
the longitudinal and transverse relative accelerations, conforming to
traditional names for these quantities. 

For the Lie case the derivative of the unit relative velocity is not
orthogonal to the velocity vector 
\begin{equation}
    \hat\nu(U,u)_\alpha 
    D_{\rm(lie)}(U,u) \hat\nu(U,u)^\alpha / d\ell_{(U,u)}
     = - 2 \theta(u)_{\alpha \beta} \hat\nu(U,u)^\alpha \hat\nu(U,u)^\beta
\end{equation}
unless the expansion tensor of the observer congruence vanishes (in
which case the Lie and corotating Fermi-Walker derivatives of the
various types agree) or unless the relative motion is along a
direction in which the observer expansion is zero. Thus one must
actually project this derivative in order to accomplish the
``direction-of-relative-motion" orthogonal decomposition. 

The transverse relative acceleration for the ordinary and corotating
Fermi-Wal\-ker cases 
\begin{eqnarray}
  a_{\rm(tem)}^{(\bot)}(U,u)^\alpha 
   &=& \nu(U,u) D_{\rm(tem)}(U,u) \hat\nu(U,u)^\alpha / d \tau_{(U,u)}
                \nonumber\\
  &=& \nu(U,u)^2 D_{\rm(tem)}(U,u) \hat\nu(U,u)^\alpha / d \ell_{(U,u)}
      \ ,
\end{eqnarray}
where Eq.~(\ref{eq:elltau}) has been used to re-parametrize the
derivative of the unit velocity vector, is exactly what one calls the
centripetal acceleration in the case of the usual inertial observers
in Minkowski spacetime, so it is natural to call it the ``relative
centripetal acceleration." 

One may next decompose the rate of change of spatial momentum in the
same way 
\begin{eqnarray}\label{eq:rca}
     D_{\rm(tem)}(U,u) p(U,u)^\alpha / d\tau_{(U,u)} 
    &=&  \hat\nu^\alpha d p(U,u) / d\tau_{(U,u)} 
  \nonumber\\ & &\quad
     + \gamma(U,u) \nu(U,u) D_{\rm(tem)}(U,u) 
                              \hat\nu(U,u)^\alpha / d\tau_{(U,u)} 
  \nonumber\\
    &=&  \hat\nu^\alpha d p(U,u) / d\tau_{(U,u)} 
    + \gamma(U,u)     a_{\rm(tem)}^{(\bot)}(U,u)^\alpha \ ,
\end{eqnarray}
where the second equality only holds for the ordinary and corotating
Fermi-Walker cases. The second term is proportional to the relative
acceleration for those cases. The first term is parallel to the
direction of motion and itself contains both the longitudinal relative
acceleration scalar $d \nu(U,u) / d\tau_{(U,u)}$ as well as the effect
of the changing three-energy (per unit mass) of the test particle when
the derivative is expanded by the product rule. 

Consider the derivatives of the unit velocity vector for the two
Fermi-Walker cases, which are related to each other by 
\begin{eqnarray}
       D_{\rm(cfw)}(U,u) \hat\nu(U,u)^\alpha / d \ell_{(U,u)} 
     &=&   D_{\rm(fw)}(U,u) \hat\nu(U,u)^\alpha / d \ell_{(U,u)} 
         \nonumber\\ &&\quad
     - \nu(U,u)^{-1} [ \vec\omega(u) \times_u \hat\nu(U,u) ]^\alpha \ .
\end{eqnarray}
If each were the uniquely defined intrinsic derivative with respect to
the arclength of the unit tangent to a curve in a three-dimensional
Riemannian manifold, the unit vector $\eta_{\rm(tem)}(U,u)^\alpha$
specifying its direction (the relative unit normal) would be the first
 normal to the curve and its magnitude $\kappa_{\rm (tem)}(U,u) \ge0$
(the relative curvature) would be the curvature of that curve, the
reciprocal of which would define a radius of curvature (the relative
radius of curvature) $\rho_{\rm (tem)}(U,u) = 1/\kappa_{\rm
(tem)}(U,u)$ when the curvature is nonzero. This leads to the
representation 
\begin{equation}
       D_{\rm(tem)}(U,u) \hat\nu(U,u)^\alpha / d \ell_{(U,u)} 
       = 1/\rho_{\rm (tem)}(U,u) \, \eta_{\rm (tem)}(U,u)^\alpha
\end{equation}
of the unit velocity derivative and 
\begin{equation}
  a^{(\bot)}_{\rm(tem)}(U,u) = \nu(U,u)^2/ \rho_{\rm (tem)}(U,u)
\end{equation}
for the magnitude of the relative centripetal acceleration, which
takes its familiar form in terms of the relative radius of curvature.
Eq.~(\ref{eq:aaa}) shows that the two accelerations differ by a term
orthogonal to the relative direction of motion. As noted above, in the
stationary case these concepts reduce to the analogous quantities in
the Riemannian 3-manifold of the observer quotient space. For an
arbitrary spacetime these generalizations can be studied to understand
the sense in which they generalize the more familiar concepts, but for
now they will be taken merely as formal definitions. 

For any test particle trajectory, the relative centripetal
acceleration is zero when the relative curvature
$\kappa_{\rm(tem)}(U,u)$ vanishes, corresponding to the limit of
infinite radius of curvature. This enables one to define trajectories
for which the curvature vanishes identically as ``relatively straight"
with respect to the ordinary or corotating Fermi-Walker total spatial
covariant derivative. Relative motion for which this is true may be
called the case of purely linear relative acceleration (for each
type), examined recently in the static case by Rindler and Mishra
\cite{rinmis,mis} and the present authors \cite{rok}. On the other
hand, if the longitudinal relative acceleration vanishes, as it does
for the case of constant relative speed in the ordinary and corotating
cases, one has the case of purely transverse relative acceleration,
the case studied extensively for circular orbits by Abramowicz et al
in the static case \cite{acl88,a90a,a90b,a90c,ab91,a92,ams93} and by
Abramowicz and coworkers in the stationary Kerr spacetime and
stationary axisymmetric spacetimes \cite{pc90,ip93,a93,anw93,anw95}.
De Felice has studied this same case for Schwarzschild and Kerr
without decomposing the 4-force and using the angular velocity
relative to the static (or distantly nonrotating) observers as his key
variable \cite{def91,defuss91,defuss93,def94,def95}. Barrab\`es,
Boisseau, and Israel \cite{barboiisr} have done the same, but using
the locally nonrotating observer relative velocity as the key
variable. 

It is worth noting that in the case of orthogonality of the vorticity
vector and the relative velocity as occurs for the circular orbits the
vanishing of the corotating Fermi-Walker relative curvature implies
that the angular velocity-like quantity $\nu(U,u)/\rho_{\rm(fw)}(u)$
of the center of relative curvature in the local rest space equals the
magnitude of the vorticity. 

\section{Spatial gravitational forces}

The spatial projection of the four-acceleration 
\begin{equation}\label{eq:aU}
        a(U)^\alpha = D U^\alpha / d \tau_U \ ,
\end{equation}
when rescaled to take into account the differences in proper times, is
the apparent three-acceleration as seen by the test observers 
\begin{eqnarray}
  A(U,u)^\alpha
   &=& \gamma(U,u)^{-1} P(u)^\alpha{}_\beta D U^\beta / d\tau_U 
  \nonumber\\
   &=& D_{\rm(fw)}(U,u) 
         [ \gamma(U,u) u^\alpha + p(U,u)^\alpha ]/ d\tau_{(U,u)}
   \nonumber\\
   &=& D_{\rm(fw)}(U,u) p(U,u)^\alpha / d\tau_{(U,u)} 
                           - F{}_{\rm(fw)}^{\rm(G)}(U,u)^\alpha \ ,
\end{eqnarray}  
and it can be rewritten in terms of the other two total spatial
covariant derivatives in a single form 
\begin{equation}
   A(U,u)^\alpha
      = D_{\rm(tem)}(U,u) p(U,u)^\alpha / d\tau_{(U,u)} 
                - F{}_{\rm(tem)}^{\rm(G)}(U,u)^\alpha \ 
\end{equation}
where ${\rm tem} = {\rm fw, cfw, lie}$.

The spatial gravitational force (per unit mass)
\begin{eqnarray}\label{eq:Ftem}
     F{}_{\rm(tem)}^{\rm(G)}(U,u)^\alpha 
      &=& \gamma(U,u) D_{\rm(tem)}(U,u) u^\alpha / d\tau_{(U,u)} 
           \nonumber\\
      &=& \gamma(U,u) [ g(u)^\alpha 
          + H_{(\rm tem)}(u)^\alpha{}_\beta \nu(U,u)^\beta] \ , 
\end{eqnarray}
is a Lorentz-like force determined by the gravitoelectric
$g(u)^\alpha$ and gravitomagnetic $H_{\rm(tem)}(u)^\alpha{}_\beta$
fields which in turn are simply related to the kinematical fields of
the observer congruence 
\begin{eqnarray}
  g(u)^\alpha &=& - a(u)^\alpha \ ,
     \nonumber\\
  H_{(\rm fw)}(u)^\alpha{}_\beta 
       &=& \omega(u)^\alpha{}_\beta -\theta(u)^\alpha{}_\beta\ ,
     \nonumber\\
  H_{(\rm cfw)}(u)^\alpha{}_\beta 
       &=& 2\omega(u)^\alpha{}_\beta -\theta(u)^\alpha{}_\beta\ ,
     \nonumber\\
  H_{(\rm lie)}(u)^\alpha{}_\beta 
       &=& 2\omega(u)^\alpha{}_\beta -2\theta(u)^\alpha{}_\beta\ .
\end{eqnarray}
It is useful to introduce a single gravitomagnetic vector field which
determines the antisymmetric part of all of the various
gravitomagnetic tensor fields 
\begin{equation}
     H(u)^\alpha = 2 \omega(u)^\alpha \ ,
\end{equation}
namely so that
\begin{equation}
    2 \omega(u)^\alpha{}_\beta \nu(U,u)^\beta 
         = [\nu(U,u) \times_u \vec H(u)]^\alpha \ .
\end{equation}
The symmetric part of the gravitomagnetic tensor field is just a
multiple of the expansion tensor for each case. 

If the acceleration of the test particle equals a spacetime force
$a(U)^\alpha = f(U)^\alpha$ and one introduces the rescaled spatial
projection 
\begin{equation}\label{eq:FPf}
   F(U,u)^\alpha 
   = \gamma(U,u)^{-1} P(u)^\alpha{}_\beta f(U)^\beta \ ,
\end{equation}
then the spatial projection of the force equation (``spatial equation
of motion") may be written in the form 
\begin{equation}
   D_{\rm(tem)}(U,u) p(U,u)^\alpha / d\tau_{(U,u)} 
        =  F^{(\rm G)}_{(\rm tem)} (U,u)^\alpha + F(U,u)^\alpha\ .
\end{equation}
The spatial gravitational force represents the combined inertial
forces due to the motion of the family of test observers. It arises in
the same way as the noninertial forces in nonrelativistic mechanics,
namely as a part of the total acceleration which is moved to the
opposite side of the ``acceleration equals force per unit mass"
equation with a sign change. 

Space curvature effects are encoded in the total spatial covariant
derivative itself. Suppose one considers a world line segment which
starts and ends on a single observer world line, and one transports a
spatial vector along both paths (general world line and observer world
line) from the initial to the final point using the transport
associated with one of the three kinds of total spatial covariant
derivatives. In each case the two final vectors will have distinct
directions, and in the Lie case, different magnitudes in general, due
to curvature effects associated with the spatial metric, similar to
the case of two such paths in the simpler case of a fixed Riemannian
three-manifold. For a timelike Killing vector field test observer
congruence in a stationary spacetime, where the Lie and corotating
total spatial covariant derivatives coincide, the corotating
Fermi-Walker space curvature effect is exactly that due to the
curvature of the natural projected Riemannian metric on the quotient
space of observer world lines. For circular orbits in a stationary
axisymmetric spacetime, this effect may be calculated explicitly 
using the tangent cone to the embedding of the plane of the orbit, 
as discussed in appendix 1.A of Arnold \cite{arn}. 

Since index shifting does not commute with the Lie total spatial
covariant derivative, letting this derivative act instead on
$p(U,u)_\alpha$ leads to an additional expansion term in the Lie
spatial gravitational force. Conveniently introducing a ``flattened"
Lie total spatial covariant derivative by 
\begin{equation}
    D_{\rm(lie\flat)}(U,u) X^\alpha / d\lambda 
           = P(u)^{\alpha\beta}
             D_{\rm(lie\flat)}(U,u) X_\beta / d\lambda \ ,
\end{equation}
and a ``flattened" Lie spatial gravitational force with a
corresponding gravitomagnetic field 
\begin{equation}
  H_{\rm(lie\flat)}(u)^\alpha{}_\beta 
       = 2\omega(u)^\alpha{}_\beta\ ,
\end{equation}
one has the analogous form of the force equation
\begin{equation}
   D_{\rm(lie\flat)}(U,u) p(U,u)^\alpha / d\tau_{(U,u)} 
        =  F^{(\rm G)}_{\rm(lie\flat)} (U,u)^\alpha + F(U,u)^\alpha\ .
\end{equation}
This notation facilitates the comparison of the different choices
without requiring index shifting. 

\section{Massless test particles}

Consider a massless test particle following a null path with affine
parameter $\lambda_P$ and tangent $P^\alpha$ locally expressible as $
dx^\alpha / d \lambda_P $ in terms of local coordinates. Interpreting
$P^\alpha$ as the 4-momentum directly, rather than the 4-momentum per
unit mass  of the previous discussion for a massive test particle, one
may essentially make the substitution 
$ (U^\alpha, \tau_U, \gamma(U,u), f^\alpha, F^\alpha) \to
(P^\alpha, \lambda_P, E(P,u), f^\alpha, F^\alpha) $
in that discussion to reinterpret the results and formulas in the new
context. Since the speed now satisfies 
\begin{equation}
     \nu(U,u) =  d \ell_{(U,u)} / d \tau_{(U,u)} = 1 \ ,
\end{equation}
the relative velocity is a unit vector $\hat\nu(P,u)^\alpha =
p(P,u)^\alpha / E(P,u)$, while the energy $E(P,u)$ and magnitude of
the spatial momentum $p(P,u)^\alpha$ are now related by $E(P,u) =
p(P,u)$. The spatial equation of motion then becomes simply 
\begin{equation}
   D_{\rm(tem)}(U,u) p(U,u)^\alpha / d\tau_{(U,u)} 
        =  F^{(\rm G)}_{(\rm tem)}(U,u)^\alpha + F(P,u)^\alpha \ ,
\end{equation}
where the spatial gravitational force is
\begin{equation}
     F^{(\rm G)}_{(\rm tem)}(U,u)^\alpha
        = E(P,u) [ g(u)^\alpha 
                + H_{(\rm tem)}(u)^\alpha{}_\beta \nu(U,u)^\beta] \ .
\end{equation}
For null geodesics the (affine-parameter-dependent) forces
$f(P)^\alpha$ and $F(P,u)^\alpha$ are zero, but accelerated photon
motion is important in the discussion of certain relativistic
phenomena like the Sagnac effect, where photons traveling in opposite
directions around a loop via mirrors or fiber optics are indeed
accelerated. 

For a massless test particle,  the ordinary and corotating
Fermi-Walker tangential relative accelerations are automatically zero
since the relative velocity is a unit vector, and the relative
centripetal acceleration has exactly the same form as for the nonzero
mass case 
\begin{equation}
  a^{(\bot)}_{\rm(tem)}(P,u) = \nu(P,u)^2/ \rho_{\rm (tem)}(P,u) 
            = 1 / \rho_{\rm (tem)}(P,u) \ ,
              \quad {\rm\scriptstyle tem = fw,\, cfw} \ .
\end{equation}
This acceleration vanishes for null trajectories which undergo
``relatively straight" relative motion. 

\section{Observer-adapted spatial frames}

The projection formalism is greatly simplified if expressed in terms
of a spacetime frame adapted to the splitting of each tangent space
defined by the test observer family. Let $\{E_a{}^\alpha\}$ be a
spatial frame, i.e., such that it is a basis of each local rest space
$LRS_u$. It is convenient to express the above results in terms of the
observer-adapted spacetime frame $\{u^\alpha,E_a{}^\alpha\}$, with
dual frame $\{ -u_\alpha, W^a{}_\alpha\}$. Spatial fields then only
have spatially-indexed frame components nonzero, like the spatial
metric $h_{ab} = P(u)_{ab}$. Note that observer-adapted spatial frame
components are distinct from the Latin-indexed coordinate components;
unless otherwise indicated Latin indices in formulas will refer to the
frame components. 

Let the frame derivatives of functions be denoted by the comma
notation 
\begin{equation}
     u(f) 
    = f_{,0} \ , \quad E_a{}^\alpha \partial_\alpha f = f_{,a} \ .
\end{equation}
To express derivatives of tensor fields, one needs the temporal and
spatial derivatives of the spatial frame vectors themselves, as well
as their Lie brackets 
\begin{eqnarray}
     \nabla_{\rm(tem)}(u) E_a{}^\alpha 
           &=& C_{\rm(tem)}(u)^b{}_a E_b{}^\alpha\ ,
        \qquad
       \nabla(u)_{\textstyle E_a} E_b{}^\alpha
           = \Gamma(u)^c{}_{ab} E_c{}^\alpha\ ,
           \nonumber\\
     \left( P(u) [ E_a,E_b ] \right)^\alpha 
         &=& C(u)^c{}_{ab} E_c{}^\alpha\ ,
\end{eqnarray}
where
\begin{equation}\label{eq:CCC}
     C_{\rm(cfw)}(u)^a{}_b  
   = C_{\rm(fw)}(u)^a{}_b + \omega(u)^a{}_b
   = C_{\rm(lie)}(u)^a{}_b + \theta(u)^a{}_b \ .
\end{equation}
For example, if $X^\alpha = X^a E_a{}^\alpha$ is a spatial vector
field, one has 
\begin{equation}
     \nabla_{\rm(tem)}(u) X^a 
        = X^a{}_{,0} + C_{\rm(tem)}(u)^a{}_b X^b 
             \ ,\quad
     \nabla(u)_b X^a = X^a{}_{,b} + \Gamma(u)^a{}_{bc} X^c \ ,
\end{equation}
while if it is only defined along the world line, one has
\begin{equation}
    D_{\rm(tem)}(U,u) X^a / d \tau_{(U,u)}
       = d X^a / d \tau_{(U,u)} + C_{\rm(tem)}(u)^a{}_b X^b
                 + \Gamma(u)^a{}_{bc} \nu(U,u)^b X^c \ .
\end{equation}

The frame components of the spatial connection are easily expressed in
terms of the spatial metric derivatives and the spatial structure
functions of the spatial frame 
\begin{equation}
    \Gamma(u)_{abc} 
      = {\textstyle \frac12} ( h_{\{ab,c\}_-} + C(u)_{\{abc\}_-})\ ,
\end{equation}
where $A_{\{abc\}_-} = A_{abc} - A_{bca} + A_{cab}$.

For an orthonormal frame, the Fermi-Walker and corotating Fermi-Walker
frame coefficients $C_{\rm(tem)}(u)^a{}_b$ are antisymmetric. For the
special Serret-Frenet orthonormal spatial frame associated with the
observer congruence, only two independent such coefficients exist; in
the Fermi-Walker case, since they arise from the spatial projection of
the Fermi-Walker derivatives of the frame vectors along the observer
world lines, they are just the first and second torsions of those
world lines \cite{iyevis93}. (Note that according to the definition
\cite{iyevis93}, a single trajectory of a nontrivial quasi-Killing
vector field is a Killing vector field trajectory, but the family of
such trajectories for a single quasi-Killing vector field consists of
trajectories not of one single Killing vector field.) 

\section{Spatial gravito-potentials}

The discussion of the measurement by a congruence of test observers of
tensor fields and of tensor differential equations using no other
spacetime structure may be referred to as the congruence point of
view. (Original references for spacetime splittings are given
elsewhere \cite{mfg}.) While in a generalized sense the 4-velocity
$u^\alpha$ serves as a 4-vector potential for the gravitoelectric and
gravitomagnetic vector force fields in this point of view (a partial
splitting of spacetime), scalar and spatial vector potentials
analogous to those in electromagnetism may be defined only by
introducing certain equivalence classes of local coordinates which are
adapted to the congruence of observer world lines in some way (a full
splitting of spacetime). For a general congruence with nonzero
vorticity, the appropriate coordinates are comoving, leading to an
approach called the threading point of view, which merely represents
the general discussion of observer measured quantities in an adapted
coordinate system. For a special case of a vorticity-free congruence
where the gravitomagnetic vector field vanishes, it is more natural to
refer to the partial splitting as the hypersurface point of view, and
for the full splitting one can use more general coordinates adapted to
the family of orthogonal hypersurfaces admitted by the congruence (the
time coordinate hypersurfaces), allowing the time coordinate lines to
be determined by a second independent congruence. One then has the
choice of representing all the hypersurface-forming observer-measured
quantities directly (the hypersurface point of view) or of working in
a hybrid two-congruence approach called the slicing point of view, in
which the evolution is described in terms of the second congruence.
This latter approach is the one well known from the work of Arnowit,
Deser, and Misner \cite{adm,mtw}. The gravitomagnetic vector field
reappears in this latter approach as a relative velocity effect
introduced through the use of a new temporal derivative along the time
lines rather than along the observer world lines \cite{thoetal}. Thus
for a given spacelike slicing and timelike threading (together forming
a ``nonlinear reference frame"), one obtains two distinct families of
observers and three distinct points of view which agree only when the
slicing and threading are orthogonal. 

Let $\{x^\alpha\} = \{t,x^a\}$ with $x^0=t$ be a set of local
coordinates (said to be ``adapted to the nonlinear reference frame")
for which the time coordinate hypersurfaces (constant $t$) belong to
the given slicing of spacetime and the time coordinate lines (constant
$x^a$) belong to the given threading congruence. Introduce also the
vector field tangent to the time coordinate lines $e_0{}^\alpha =
\delta^\alpha{}_0$. 

\subsection{The threading point of view}

First consider the threading point of view, which is especially useful
in a stationary spacetime where it enables one to interpret the
spacetime geometry in terms of the quotient space geometry on the
space of Killing observers. The adapted coordinates are comoving
coordinates for the 4-velocity of the observer congruence $u^\alpha =
m^\alpha = M^{-1} \delta^\alpha{}_0$, and the observer world lines
coincide with the time coordinate lines. 

The spacetime line element (covariant metric) takes the form 
\begin{eqnarray}
    ds^2 &=&    -M^2(dt -M_a dx^a)^2 +\gamma_{ab}dx^a dx^b \ ,
                     \nonumber\\
         &=&    M^2[ -(dt -M_a dx^a)^2 +\tilde\gamma_{ab}dx^a dx^b ]
                   \ ,
\end{eqnarray}
where $\gamma_{ab}$ parametrizes the spatial metric and is the matrix
inverse of $\gamma^{ab} = P(m)^{ab}$, while $       \tilde\gamma_{ab}
= M^{-2} \gamma_{ab} $ parametrizes the ``optical spatial metric"
$\tilde P(m)_{\alpha\beta} = M^{-2} P(m)_{\alpha\beta}$ obtained by a
conformal rescaling of the spatial metric, using the terminology of
Abramowicz et al \cite{acl88}. The ``spatial derivatives" 
$\epsilon_a = \partial / \partial x^a + M_a \partial / \partial t
= \epsilon_a{}^\alpha \partial / \partial x^\alpha$
define the basis vector fields $\{\epsilon_a{}^\alpha\}$ of the local
rest space of the test observers (which together with $m^\alpha$ form
an observer-adapted frame for which 
\hfil\break
\typeout{*** Forced line break here because of math symbol in margin.}
$C_{\rm(lie)}(m)^a{}_b = 0$ and
$C(m)^a{}_{bc} = 0$), and the matrix $\gamma_{ab}$ is the matrix of
their inner products. The associated components of the spatial
connection 
\begin{equation}
     \epsilon_a{}^\alpha \nabla_\alpha(m) \epsilon_b{}^\beta
      = \Gamma(m)^c{}_{ab} \epsilon_c{}^\beta \ ,
\end{equation}
expressable as
\begin{equation}
     \Gamma(m)^c{}_{ab}
     = {\textstyle {1\over2}} \gamma^{cd}( \gamma_{da,b} 
                             - \gamma_{ab,d} + \gamma_{bd,a} ) \ ,
\end{equation}
where $f_{,a} = \epsilon_a{}^\alpha \partial f / \partial x^\alpha$,
may be used to evaluate the spatial covariant derivative of a spatial
vector field $X^\alpha = X^a \epsilon_a{}^\alpha$ (parametrized by its
spatially-indexed contravariant coordinate components $X^a$) entirely
in terms of the components in that frame in the usual way \cite{mfg}.
Analogous tilde expressions hold for the components
$\tilde\Gamma(m)^c{}_{ab}$ of the optical connection
$\tilde\nabla(m)_\alpha$. 

The shift 1-form $M_\alpha = M_a \delta^a{}_\alpha$ is a spatial
1-form which determines the shift of the orientation of the local rest
spaces of the test observers away from the coordinate time
hypersurfaces, while the lapse function $M$ relates coordinate time
along the time lines to the test observer proper time. The combination
$\nu(n,m)^\alpha = M M^a \delta^\alpha{}_a$ is the relative velocity
field of the normal trajectories to the coordinate time hypersurfaces.
The lapse and shift serve as scalar and vector potentials for the
gravitoelectric and gravitomagnetic vector fields of the test observer
congruence, while the spatial metric generates the symmetric part of
the gravitomagnetic tensor field, which is proportional to the
expansion tensor 
\begin{eqnarray}
    g(m)_\alpha &=& - a(m)_\alpha=  - \nabla(m)_\alpha \ln M 
              - \pounds(m)_{\hbox{$e_0$}} M_\alpha \ ,
              \nonumber\\
    H(m)^\alpha &=&  2 \omega(m)^\alpha 
           = M \eta(m)^{\alpha\beta\gamma} \nabla(m)_\beta M_\gamma
           = M [\nabla(m) \times_m \vec M]^\alpha \ ,\\
    \theta(m)_{\alpha\beta} 
           &=& {\textstyle {1\over2}} 
               \pounds(m)_{\hbox{$e_0$}} P(m)_{\alpha\beta}
                \ .\nonumber
\end{eqnarray}
In the observer-adapted frame only the spatially indexed components
(distinct from the coordinate components of this type) of these fields
are nonzero, and the Lie derivatives reduce to the partial derivatives
of these components with respect to the $t$ coordinate, for example 
\begin{equation}
  \theta(m)_{ab} = {\textstyle {1\over2}} \partial_t \gamma_{ab} \ .
\end{equation}
The total spatial covariant derivative of the spatial momentum becomes
explicitly 
\begin{eqnarray}
     && D_{\rm(tem)}(U,m) p(U,m)^a / d \tau_{(U,m)}
\\
     &&  = d p(U,m)^a / d \tau_{(U,m)} 
            + \gamma(U,m) C_{\rm(tem)}(m)^a{}_b \nu(U,m)^b
            - F^{\rm(SC)}(U,m)^a\ ,          
              \nonumber
\end{eqnarray}
where 
\begin{equation}
  F^{\rm(SC)}(U,m)^a 
       = - \gamma(U,m)\Gamma(m)^a{}_{bc} \nu(U,m)^b \nu(U,m)^c
\end{equation}
defines the ``space curvature" force in the threading point of view. 

\subsection{The hypersurface and slicing points of view}

In the hypersurface and slicing points of view, given a family of
spacelike hypersurfaces with unit normal $n^\alpha$, the spacetime
line element in adapted local coordinates takes the form 
\begin{eqnarray}
   ds^2 &=& -N^2 dt^2 + g_{ab}(dx^a + N^a dt)(dx^b +N^b dt)  
                 \nonumber \\
        &=& N^2[-dt^2 + \tilde g_{ab}(dx^a + N^a dt)(dx^b +N^b dt)] 
                 \ .
\end{eqnarray}
Here $g_{ab}= P(n)_{ab}$ parametrizes the spatial metric and is the
matrix inverse of $ g^{ab} $, while $       \tilde g_{ab} = N^{-2}
g_{ab} $ parametrizes the ``optical spatial metric" $\tilde
P(n)_{\alpha\beta} = N^{-2} P(n)_{\alpha\beta}$. The spatial
coordinate vectors themselves $ e_a{}^\alpha = \delta^\alpha{}_a $
form a spatial frame (which together with $n^\alpha$ form an
observer-adapted frame with $C(n)^a{}_{bc}=0$ and
$C_{\rm(lie)}(n)^a{}_b = N^{-1}\partial N^a /\partial x^b$) which may
be used to express the spatial covariant derivative of a spatial
vector field $X^\alpha = X_b g^{ba}e_a{}^\alpha$ (parametrized by its
spatially-indexed covariant coordinate components $X_a$) in terms of
the associated connection components, namely 
\begin{equation}
     e_a{}^\alpha \nabla_\alpha(n) e_b{}^\beta
      = \Gamma(n)^c{}_{ab} e_c{}^\gamma \ ,
\end{equation}
expressable as
\begin{equation}
     \Gamma(n)^c{}_{ab}
     = {\textstyle {1\over2}} g^{cd}( g_{da,b} - g_{ab,d} + g_{bd,a} ) \ ,
\end{equation}
where here $f_{,a} = \partial f / \partial x^a$. Analogous tilde
expressions hold for the components $\tilde\Gamma(n)^c{}_{ab}$ of the
optical connection $\tilde\nabla(n)_\alpha$. 

The 4-velocity of the associated test observers is the unit normal
\begin{equation}
      n^\alpha = N^{-1} [ \delta^\alpha{}_0
                             - N^a \delta^\alpha{}_a ] \ ,
\end{equation}
and their world lines are the orthogonal trajectories to the time
hypersurfaces. The shift vector field $N^\alpha = N^a
\delta^\alpha{}_a$ is a spatial vector field which determines the
shift of the time lines away from these orthogonal trajectories, while
the lapse function $N$ relates the coordinate time along the observer
world lines to the observer proper time. The combination 
$\nu(e_0,n)^\alpha = N^{-1}N^a \delta^\alpha{}_a$ is the relative 
velocity field of the threading curves.

The distinguishing feature of the slicing point of view compared to
the hypersurface point of view or the threading point of view for the
same congruence of observers (the latter requiring a comoving
coordinate system) is that it uses  a new Lie temporal derivative
along the threading curves rather than along the observer world lines 
\begin{equation}
      \nabla_{\rm(lie)}(n,e_0)= N^{-1} \pounds(n)_{\hbox{$e_0$}} 
     = \nabla_{\rm(lie)}(n)  + N^{-1} \pounds(n)_{\hbox{$\vec N$}}
              \ .
\end{equation}
This in turn leads to a new Lie spatial total covariant derivative
along a test particle worldline 
\begin{eqnarray}
       D_{\rm(lie)}(U,n,e_0) X^\alpha / d \tau_{(U,n)} 
         &=& [ \nabla_{\rm(lie)}(n,e_0)
                 + \nu(U,n)^\beta \nabla(n)_\beta ] X^\alpha 
\\
         &=& D_{\rm(lie)}(U,n) X^\alpha / d \tau_{(U,n)} 
                         - [N^{-1}  \nabla(n)_\beta N^\alpha] X^\beta \ .
          \nonumber
\end{eqnarray}
Expressing the two Lie total covariant derivatives in the
observer-adapted frame leads to 
\begin{eqnarray}
    D_{\rm(lie)}(U,n) X^a / d \tau_{(U,n)}
       &=&  d X^a / d \tau_{(U,n)} 
                 + \Gamma(n)^a{}_{bc} \nu(U,n)^b X^c 
                + X^b N^{-1} \partial N^a/\partial x^b 
                      \nonumber\\
       &=&  d X^a / d \tau_{(U,n)} 
                 + \Gamma(n)^a{}_{bc} [ \nu(U,n)^b - \nu(e_0,n)^b ] X^c 
                      \nonumber\\ &&\quad
                + X^b N^{-1} \nabla(n)_b N^a 
                      \nonumber\\
       &=&   D_{\rm(lie)}(U,n,e_0) X^a / d \tau_{(U,n)}
                            + X^b N^{-1} \nabla(n)_b N^a \ .
\end{eqnarray}
When $X^a = \nu(U,n)^a$, these equations define the slicing Lie
relative acceleration $a_{\rm(lie)}(U,n,e_0)^\alpha$ and its relation
to the hypersurface quantity, namely 
\begin{equation}
    a_{\rm(lie)}(U,n,e_0)^a - a_{\rm(lie)}(U,n)^a
        = - \nu(U,n)^b N^{-1} \nabla(n)_b N^a \ .
\end{equation}

The spatial equation of motion of a test particle then takes the form
\begin{equation}
       D_{\rm(lie)}(U,n) p(U,n)^\alpha / d \tau_{(U,n)}
           = \gamma(U,n) [ g(n)^\alpha 
                    + H_{\rm(lie)}(n)^\alpha{}_\beta \nu(U,n)^\beta ]
                    + F(U,n)_\alpha
\end{equation}
in the hypersurface point of view and
\begin{eqnarray}
       D_{\rm(lie)}(U,n,e_0) p(U,n)^\alpha / d \tau_{(U,n)}
           &=& \gamma(U,n) [ g(n)^\alpha 
                + H_{\rm(lie)}(n,e_0)^\alpha{}_\beta \nu(U,n)^\beta ]
                  \nonumber\\ &&\quad
                + F(U,n)^\alpha
                \ ,\nonumber\\
       D_{\rm(lie)}(U,n,e_0) p(U,n)_\alpha / d \tau_{(U,n)} 
           &=& \gamma(U,n) [ g(n)_\alpha 
            + H_{\rm(lie\flat)}(n,e_0)_{\alpha\beta} \nu(U,n)^\beta ]
                  \nonumber\\ &&\quad
                + F(U,n)_\alpha
\end{eqnarray}
in the slicing point of view, where the hypersurface Lie
gravitomagnetic tensor field is just minus twice the expansion tensor
when the equations of motion are expressed in contravariant form, but
zero in the covariant form 
\begin{eqnarray}
     H_{\rm(lie)}(n)_{\alpha\beta} 
          &=&  N^{-1} [ 2\nabla(n)_{(\alpha} N_{\beta)}
               - \pounds(n)_{\hbox{$e_0$}} P(n)_{\alpha\beta} ]
           = - 2 \theta(n)_{\alpha\beta}\ ,
                    \nonumber\\
     H_{\rm(lie\flat)}(n)_{\alpha\beta} &=& 0
\end{eqnarray}
and the slicing gravitomagnetic tensor fields are
\begin{eqnarray}
     H_{\rm(lie)}(n,e_0)_{\alpha\beta} 
          &=&  N^{-1} [ \nabla(n)_\alpha N_\beta
                  - \pounds(n)_{\hbox{$e_0$}} P(n)_{\alpha\beta} ]
             \ ,\nonumber\\
     H_{\rm(lie\flat)}(n,e_0)_{\alpha\beta} 
          &=&  N^{-1}  \nabla(n)_\alpha N_\beta \ .
\end{eqnarray}
The contravariant and covariant forms of the equation of motion differ
by a term arising from the Lie temporal derivative of the spatial
metric. For the contravariant form of the equation of motion the
symmetric part of the gravitomagnetic tensor differs from $-2
\theta(n)_{\alpha\beta}$ by the missing reversed-index shift
derivative term which would symmetrize the term which is present. This
missing term now contributes by its absense to the slicing
gravitomagnetic vector field but is restored by an extra shift
derivative term in the corresponding second order acceleration
equation due to the relative velocity of the observers and evolution
curves, thus restoring the analogy with the threading point of view at
that level \cite{mfg}. 

The lapse and shift serve as scalar and vector potentials for the
gravitoelectric and gravitomagnetic vector fields in the slicing point
of view. The gravitoelectric field is still the sign-reversed
acceleration of the test observers in both points of view, but the
slicing gravitomagnetic vector field is now a result of the relative
motion of the time lines with respect to the observer world lines 
\begin{eqnarray}
    g(n)_\alpha &=& - a(n)_\alpha=  - \nabla(n)_\alpha \ln N \ ,
 \qquad
    H_{\rm(lie)}(n)^\alpha = 0 \ ,
                         \nonumber\\
    H_{\rm(lie)}(n,e_0)^\alpha   
    &=& N^{-1} \eta(n)^{\alpha\beta\gamma} \nabla(n)_\beta N_\gamma
                = N^{-1} [\nabla(n) \times_n \vec N]^\alpha \ ,
\end{eqnarray}
In terms of the gravitomagnetic vector field, the slicing spatial
equation of motion takes the form 
\begin{eqnarray}
   D_{\rm(lie)}(U,n,e_0) p(U,n)^\alpha / d \tau_n 
       &=& \gamma(U,n) \{ g(n)^\alpha 
        + {\textstyle {1\over2}} 
          [ \nu(U,n) \times_n \vec H_{\rm(lie)}(n,e_0) ]^\alpha
              \nonumber\\ &&\quad
   + H_{\rm(lie)}^{\rm(SYM)}(n,e_0)^\alpha{}_\beta \nu(U,n)^\beta \} 
       \ .
\end{eqnarray}
All of these expressions have simple analogous forms when expressed in
the observer-adapted frame. The total spatial covariant derivative of
the spatial momentum 
becomes explicitly
\begin{eqnarray}
      D_{\rm(lie)}(U,n) p(U,n)^a / d \tau_{(U,n)}
       &=& d p(U,n)^a / d \tau_{(U,n)} - F^{\rm(SC)}(U,n)^a  
                \nonumber\\ &&\quad
           + p(U,n)^b \partial N^a / \partial x^b \ ,
\\
      D_{\rm(lie)}(U,n,e_0) p(U,n)^a / d \tau_{(U,n)}
       &=& d p(U,n)^a / d \tau_{(U,n)} 
             - F^{\rm(SC)}(U,n,e_0)^a  
    \nonumber
\end{eqnarray}
where
\begin{eqnarray}
  F^{\rm(SC)}(U,n)^a 
    &=& - \gamma(U,n)\Gamma(n)^a{}_{bc} \nu(U,n)^b \nu(U,n)^c
          \ ,\nonumber\\
  F^{\rm(SC)}(U,n,e_0)^a 
    &=& - \gamma(U,n)\Gamma(n)^a{}_{bc} 
                 [\nu(U,n)^b -  N^{-1} N^b]\nu(U,n)^c
\end{eqnarray}
defines the ``space curvature" force in the two points of view. 

\subsection{Stationary spacetimes}

Suppose $e_0{}^\alpha$ (and therefore $m$) is a timelike Killing
vector field in some open submanifold of a stationary spacetime,
implying that $\theta(m)_{\alpha\beta}=0$, and suppose that $n^\alpha$
is also timelike in some other open submanifold overlapping with the
first. One then has a nonlinear reference frame adapted to the
stationarity in which one may consider threading, slicing, and
hypersurface points of view. The Kerr black hole spacetimes in
Boyer-Lindquist coordinates which are adapted to the distantly
nonrotating (static) observers determining the threading congruence
and to the locally nonrotating observers determining the slicing
congruence are a good example to keep in mind. 

All Lie derivatives of stationary fields along $e_0{}^\alpha$ vanish.
This eliminates the complication of the mixing of time coordinate
derivatives with spatial coordinate derivatives in the spatial
derivatives of component fields in threading point of view when
differentiating stationary tensor fields, since their various
components are independent of the time coordinate. In all cases the
spatial metric projects to a Riemannian metric on the threading
quotient space where the relative motion takes place and all
calculations become much more straightforward. The scalar and vector
potentials then become potentials in the usual sense on this space
since spatial projection is unnecessary and the curl and gradient are
just the ones on this Riemannian observer quotient space. Furthermore
the threading observer expansion tensor vanishes although the slicing
observer one does not, leading to the coincidence of the Lie and
corotating Fermi-Walker derivatives in the threading point of view.
The temporal derivative observer-adapted frame structure functions in
the threading point of view therefore satisfy $C_{\rm(cfw)}(m)^a{}_b =
C_{\rm(lie)}(m)^a{}_b = 0$, and the two associated total spatial
covariant derivatives coincide as well. In the case of the slicing
observer-adapted frame, one instead has 
\begin{equation}
     C(n)_{\rm(cfw)}{}^a{}_b  -  \theta(n)^a{}_b 
  =  C(n)_{\rm(lie)}{}^a{}_b 
  =  N^{-1} \partial N^a / \partial x^b \ , 
\end{equation}
but for relative motion along a spatial Killing direction $ \nu(U,n)^b
\partial N^a / \partial x^b = 0$, so the shift derivative term does
not contribute to the relative curvature and the centripetal
acceleration in that case, making the Lie relative curvature the
simplest such curvature in the hypersurface point of view. However,
the expansion term does lead to an additional term in the corotating
Fermi-Walker centripetal acceleration which is linear in the relative
velocity; a similar linear term arises also in the the slicing point
of view expression due to the relative velocity of the observers
themselves relative to the threading. 

\section{Spatial coordinate line curvature in stationary spacetimes
         with additional symmetry}   

Suppose $e_0{}^\alpha$ is a timelike Killing vector field in a
stationary spacetime with an additional spacelike Killing vector field
for which $t$ and $x^3$ respectively are comoving coordinates, so that
the metric only depends on the remaining two coordinates. Assume also
that the spatial coordinates are orthogonal, a situation which
describes many interesting stationary spacetimes, including rotating
black holes (Kerr spacetimes) in Boyer-Lindquist coordinates and the
G\"odel spacetime in the usual Cartesian-like or cylindrical-like
coordinates. 

Consider a test particle worldline following a Killing trajectory in
the 2-surface of the coordinates $t$ and $x^3$ with the remaining two
coordinates fixed, implying that in either the associated threading or
slicing point of view, the unit velocity vector has a single
nonvanishing constant spatial coordinate component. Constant speed
circular orbits in the above-mentioned spacetimes are of this type,
for example. 

Consider the threading point of view, for example, where
$\hat\nu(U,m)^a = \gamma_{33}{}^{-1/2} \delta^a{}_3$ and 
\begin{equation}
   D_{\rm(lie)}(U,m) \hat\nu(U,m)^a / d \ell_{(U,m)}
   = \Gamma(m)^a{}_{33} (\hat\nu(U,m)^3)^2
   = - \gamma_{aa}{}^{-1} \partial 
               (\ln \gamma_{33}{}^{1/2}) / \partial x^a 
  \ ,
\end{equation}
or, since the spatial coordinates are orthogonal, in terms of the
associated orthonormal (``physical") components indicated with the
``hat" notation 
\begin{equation}
    D_{\rm(lie)}(U,m) \hat\nu(U,m)^{\hat a} / d \ell_{(U,m)}
     = - \gamma_{aa}{}^{-1/2} \partial 
                       (\ln \gamma_{33}{}^{1/2}) / \partial x^a
     = \kappa(3,m)^{\hat a}  \ ,
\end{equation}
which can only be nonzero for $a=1,2$. The Lie relative curvature
$\kappa_{\rm(lie)}(U,m)$ will just be the square root of the sum of
the squares of the these two physical components; in fact for circular
orbits only one of these will be nonzero. (For the case of additional
symmetry like stationary cylindrically symmetric spacetimes of which
the G\"odel spacetime is an example, one can evaluate this spatial
curvature for each spatial coordinate line along a Killing vector
field, with the result for a general timelike Killing trajectory being
obtained in a simple way from the individual results, i.e., as for a
helical Killing trajectory in G\"odel.) The corresponding optical
relative curvature has the same formula with the tilde metric and
derivative, the latter of which is described in the appendix. A
similar formula holds for the physical observer-adapted frame
components in the hypersurface point of view 
\begin{equation}
   D_{\rm(lie)}(U,n) \hat\nu(U,n)^{\hat a} / d \ell_{(U,n)}
    =  - g_{aa}{}^{-1/2} 
                 \partial (\ln g_{33}{}^{1/2}) / \partial x^a
    =  \kappa(3,n)^{\hat a}  \ ,
\end{equation}
and for its optical generalization. All formulas continue to hold for
massless test particle trajectories as well. 

These coordinate curvature quantities $\kappa(3,u)^{\hat a}$ will be
referred to as the signed Lie relative curvatures (``signed" since
they may take all real values). In each of these four formulas, the
relative curvature of these special trajectories  corresponds exactly
to the coordinate line curvature in the quotient space Riemannian
geometry or its optical version for the relevant observer point of
view (threading, slicing, or hypersurface). In the static case, all of
these points of view coincide and only two distinct relative Lie
curvatures exist. 

\section{Circular orbits in stationary axisymmetric spacetimes}

Now specialize to the case of a stationary axisymmetric spacetime with
coordinates $\{t,z,\rho,\phi\}$, where $t$ and $\phi$ are comoving
with respect to the Killing vector fields associated with the
stationary and axial symmetries respectively, and the latter three
coordinates are orthogonal, as occurs for the stationary cylindrically
symmetric G\"odel and rotating Minkowski spacetimes. Only the single
threading observer-adapted spatial frame vector
$\varepsilon_\phi{}^\alpha = \delta^\alpha{}_\phi + M_\phi
\delta^\alpha{}_t$ differs from the orthogonal spatial coordinate
frame vectors. 

For circular orbits, the observer-adapted physical components of the
relative velocity along the angular direction are 
\begin{eqnarray}
  \nu(U,m)^{\hat\phi} 
    &=& M^{-1} \gamma_{\phi\phi}{}^{1/2} \dot\phi/(1-M_\phi \dot\phi) 
       \ ,\nonumber\\
  \nu(U,n)^{\hat\phi} 
    &=& N^{-1} g_{\phi\phi}{}^{1/2} ( \dot\phi + N^\phi) \ ,
\end{eqnarray}
while
\begin{equation}
  \nu(U,n)^{\hat\phi} -  \nu(e_0,n)^{\hat\phi}
    = N^{-1} g_{\phi\phi}{}^{1/2} \dot\phi \ ,
\end{equation}
where the coordinate angular velocity $\dot\phi = d\phi/dt$ along the
test particle world line is constant for the stationary circular
orbits of constant speed to be considered here. Thus the test particle
world line is a Killing trajectory and the ordinary time derivative
terms vanish in the total spatial covariant derivatives along the test
world line. 

Attention will be confined to those circular orbits for which the
$z$-derivative relative curvature expression vanishes 
\begin{eqnarray}
  \kappa(\phi,m)^{\hat z} &=&   \kappa(\phi,n)^{\hat z} = 0
         \ ,\nonumber\\
  \kappa(\phi,m)^{\hat\rho} 
    &=&  - \gamma_{\rho\rho}{}^{-1/2} 
          \partial (\ln \gamma_{\phi\phi}^{1/2}) / \partial \rho 
        \ , 
             \nonumber\\
              \qquad
  \kappa(\phi,n)^{\hat\rho}
    &=&  - g_{\rho\rho}{}^{-1/2} 
        \partial (\ln g_{\phi\phi}^{1/2}) / \partial \rho  \ .
\end{eqnarray}
This limits one to the equatorial circular orbits in the Kerr
spacetime. The Lie relative centripetal acceleration then has the
single nonzero physical observer-adapted frame component 
\begin{equation}
     a_{\rm(lie)}(U,u)^{(\bot)}{}^{\hat\rho}
       = \kappa(\phi,u)^{\hat\rho} | \nu(U,u)^{\hat\phi} |^2 
             \ ,\quad  u = m,n \ .
\end{equation}
Since $\theta(n)_{\phi\phi} =0$, the hypersurface point of view Lie
relative acceleration is fortuitously orthogonal to the relative
velocity and so directly represents the relative centripetal
acceleration. 

The nonzero observer-adapted physical components of the various fields
are 
\begin{eqnarray}
 g(m)_{\hat\rho} &=& - (\gamma_{\rho\rho})^{-1/2} (\ln M)_{,\rho} \ ,
   \qquad
 H(m)^{\hat{z}} = M (\gamma_{\rho\rho}\gamma_{\phi\phi})^{-1/2}
	M_{\phi,\rho} 
                      \ , \nonumber\\
 g(n)_{\hat\rho} &=& - (g_{\rho\rho})^{-1/2} (\ln N)_{,\rho} 
            \ , 
 H(n,e_0)^{\hat{z}} 
       = N^{-1} (g_{\rho\rho}g_{\phi\phi})^{-1/2} N_{\phi,\rho} 
            \ ,
\end{eqnarray}
and
\begin{eqnarray}
   H(n,e_0)_{(\hat\rho\hat\phi)} 
   &=& {\textstyle {1\over2}} H(n)_{\hat\rho\hat\phi} 
    = -  \theta(n)_{\hat\rho\hat\phi}  
    = {\textstyle {1\over2}} 
          N^{-1} (g_{\phi\phi}/g_{\rho\rho})^{1/2} N^\phi{}_{,\rho}  
           \nonumber\\
    &=& {\textstyle {1\over2}} 
              C_{\rm(lie)}(n)^{\hat\phi}{}_{\hat\rho}  
     =  {\textstyle {1\over2}} 
             C_{\rm(cfw)}(n)^{\hat\phi}{}_{\hat\rho}  \ ,
\end{eqnarray}
where the comma indicates the coordinate partial derivative here.

The physical 4-force responsible for the motion of the test particle
on the circular orbit is related to the relative force by equation
(\ref{eq:FPf}). Since the projection $P(u)$ acts as the identity in
the radial direction orthogonal to the relative motion, one has the
simpler relationship between the 4-force, the 4-acceleration, and the
relative forces 
\begin{equation}
     f(U)^{\hat\rho} 
         = \gamma(U,u) F(U,u)^{\hat\rho}
         = \gamma(U,u) [ - F^{\rm(SC)}(U,u)^{\hat\rho}
                         - F^{\rm(G)}(U,u)^{\hat\rho} ]
         = a(U)^{\hat\rho} \ .
\end{equation}
The middle equality is the equation of motion for the circular orbit,
while the last equality is the value of the acceleration of the orbit.
The same considerations apply to circular orbits in the equatorial
plane of the Kerr spacetime and its Schwarzschild limit in terms of
the usual ``spherical" radial coordinate in that plane. 

Iyer and Vishveshwara have given the complete Serret-Frenet frame
formulas for the stationary axisymmetric Killing trajectories
(arbitrary constant speed circular orbits) in Kerr, G\"odel, Minkowski
and several other spacetimes \cite{iyevis93}. These depend only on the
single ratio of the two Killing vector components of the tangent
vector $\delta^\alpha{}_{t} + \omega \delta^\alpha{}_\phi$, or
equivalently of $\omega^{-1} \delta^\alpha{}_{t} +
\delta^\alpha{}_\phi$ because of its normalization. The limit
$\omega^{-1} = 0$ of their formula for the curvature $\kappa$ of the
associated Killing trajectory gives exactly
$|\kappa(\phi,n)^{\hat\rho}|$, with their frame vectors reducing to 
$(e_{(0)}^\alpha, e_{(1)}^\alpha, e_{(2)}^\alpha) 
= (-e_{\hat\phi}^\alpha, e_{\hat\rho}^\alpha, -n^\alpha)$.

\section{The Abramowicz et al Approach Demystified}

By introducing the optical metric in the threading or slicing points
of view, the full spacetime metric is conformal to one with unit lapse
and the same shift 1-form or vector field respectively, the overall
conformal factor being the square of the lapse function in each point
of view. Since null geodesics are invariant under spacetime conformal
transformations, one may re-express the spatial equation of motion for
a massless test particle in terms of the new line element, thus
absorbing the scalar potential part of the gravitoelectric force term
into the spatial geometry itself (since the rescaled metric has a unit
lapse). In the case of a static spacetime with a nonlinear reference
frame adapted to the threading congruence of a timelike Killing vector
field, so that the shift field is zero and the adapted coordinates are
Gaussian normal with respect to the new geometry, the spacetime
geodesics of the rescaled spacetime metric project down to geodesics
of the observer quotient space with the optical metric. Thus the
geodesics of the optical spatial metric on this space are the paths of
light rays, and the optical geometry measures deviations from these
paths. Of course timelike geodesics are not invariant under conformal
transformations, so the equations of motion of a massive test particle
still contain an explicit gravitoelectric term when re-expressed in
this way. 

This idea, closely related to the general relativistic Fermat
principle \cite{per} and older discussions of M\o ller \cite{mol}, is
the origin of a long series of papers by Abramowicz and coworkers. The
key difficulty in understanding their calculations associated with
this idea lies in the obscure representation both of the intrinsic
derivative along the test particle world line and of the kinematical
quantities of the observer congruence (which are not clearly
identified) in terms of nongeometrical derivatives of a vector field
on spacetime possessing the given world line as an integral curve. The
very special circumstances of circular orbits in their applications
also give false impressions of the general case; in particular the
various centripetal accelerations and gravitoelectric and
gravitomagnetic forces there are all transverse to the direction of
relative motion, and so reside in the intersection of the local rest
spaces of the observers and the test world line. The most recent
formulation of this work \cite{anw95} is simply the hypersurface point
of view description of the decomposition of the spatial projection of
the 4-acceleration of the test world line. 

The details of the spacetime conformal transformation are
straightforward and are discussed in the appendix. The new spatial
equation of motion in the threading point of view for a massless test
particle following a null geodesic turns out to be 
\begin{eqnarray}
 &&   \tilde D_{\rm(tem)}(P,m) \tilde p(P,m)_\alpha / d \tau_{(P,m)} 
                \nonumber\\ \qquad
 &&       =  M^{-1} \tilde E(P,m) 
            [ - \pounds(m)_{\hbox{$e_0$}} M_\alpha
                +  H_{(\rm tem\flat)}(m)_{\alpha\beta} 
                   \nu(P,m)^\beta ] \ ,
\end{eqnarray}
where the flat notation $\flat$ in the gravitomagnetic field is only
needed for the Lie case ${\scriptstyle {\rm tem = lie}}$, while in the
slicing point of view it is 
\begin{equation}
       \tilde D_{\rm(lie)}(P,n,e_0) \tilde p(P,n)_\alpha / d \tau_{(P,n)} 
           = N^{-1} \tilde E(P,n)
                    H_{\rm(lie\flat)}(n,e_0)_{\alpha\beta} \nu(P,n)^\beta \ .
\end{equation}
The covariant rather than contravariant form of this equation is given
for comparison with the form often found in the literature
\cite{anw95}. Here the explicit factors of the lapse correct for the
proper time and the other untransformed spatial quantities still
present. 

Thus in the case of a static spacetime with an orthogonal slicing and
threading adapted to the timelike vorticity-free Killing vector field,
the threading and slicing points of view coincide ($u=m=n$), the shift
and expansion tensor both vanish, all the various total spatial
covariant derivatives agree, and the equation of motion reduces to the
geodesic equation in the time-independent geometry of the observer
quotient space 
\begin{equation}
    \tilde D_{\rm(tem)}(P,u) \tilde p(P,u)^\alpha / d \tau_{(P,u)} 
            = 0\ .
\end{equation}
In this special case only, one can interpret the optical total spatial
covariant derivative as measuring the deviation of particle motion
from ``optically straight line paths" in the quotient space, as
advocated by Abramowicz et al \cite{acl88,a90a,a90b}. In the
stationary case additional gravitomagnetic tensor effects deflect the
null geodesics from the optical spatial geodesics in the observer
quotient spaces. 

The spatial equation of motion for massive test particles may also be
expressed in terms of the conformally rescaled quantities. Using the
results of the appendix one finds for the threading point of view 
\begin{eqnarray}
 &&   \tilde  D_{\rm(tem)} (U,m) \tilde p(U,m)_\alpha / d\tau_{(U,m)} 
          \nonumber\\
 &&\qquad
           {}= - \gamma(U,m)^{-1}  \nabla(m)_\alpha \ln M
             - \gamma(U,m) \pounds(m)_{\hbox{$e_0$}} M_\alpha 
           \nonumber\\
 &&\qquad\phantom{{}=}
    + \gamma(U,m)  H_{(\rm tem\flat)}(m)_{\alpha\beta} \nu(U,m)^\beta
              + F(U,m)_\alpha \ ,
\end{eqnarray}
where the flat subscript is only needed in the Lie case.
Correspondingly the expression for the spatial projection of the test
particle 4-acceleration (just the gamma factor times the apparent
three-acceleration) can be written 
\begin{eqnarray}
 \gamma(U,m) A(U,m)_\alpha
         &=&  \gamma(U,m) \hat\nu(U,m)_\alpha 
                  d p(U,m) / d\tau_{(U,m)} 
  \nonumber\\
        & &\ + \gamma(U,m)^2 [ \nu(U,m)^2 
          D_{\rm(tem)}(U,m) \hat\nu(U,m)_\alpha / d \ell_{(U,m)} 
  \nonumber\\
         & &\   + \nabla(m)_\alpha \ln M
                     + \pounds(m)_{\hbox{$e_0$}} M_\alpha 
  \nonumber\\
     & &\ - H_{(\rm tem\flat)}(m)_{\alpha\beta}\nu(U,m)^\beta ] \ .
\end{eqnarray} 
This in turn can be rewritten in terms of the optical metric and the
natural conformally rescaled quantities introduced in the appendix as 
\begin{eqnarray}
  \gamma(U,m) A(U,m)_\alpha
         &=& \gamma(U,m) \tilde{\hat\nu}(U,m)_\alpha 
                 d \tilde p(U,m) / d\tau_{(U,m)} 
  \nonumber\\
        & &\ +  \gamma(U,m)^2 \{ \nu(U,m)^2 
         \tilde D_{\rm(tem)}(U,m) \tilde{\hat\nu}(U,m)_\alpha 
                                           / d \tilde\ell_{(U,m)} 
  \nonumber\\
         & &\   + \gamma(U,m)^{-2}\nabla(m)_\alpha \ln M
          +  \pounds(m)_{\hbox{$e_0$}} M_\alpha
  \nonumber\\
    & &\ - H_{(\rm tem\flat)}(m)_{\alpha\beta}\nu(U,m)^\beta \} \ .
\end{eqnarray} 
Similarly, expressing the spatial projection of the 4-acceleration in
the hypersurface point of view leads to 
\begin{eqnarray} \label{eq:abram}
  \gamma(U,n) A(U,n)^\alpha
         &=&  \gamma(U,n) \tilde{\hat\nu}(U,n)^\alpha  
                  d \tilde p(U,n) / d\tau_{(U,n)} 
  \nonumber\\
        & &\ +  \gamma(U,n)^2 [ \nu(U,n)^2 
         \tilde D_{\rm(lie)}(U,n) \tilde{\hat\nu}(U,n)^\alpha
                                            / d \tilde\ell_{(U,n)} 
  \nonumber\\
         & &\   + \gamma(U,n)^{-2}\nabla(n)^\alpha \ln N
  \nonumber\\
         & &\ - H_{\rm(lie)}(n)^\alpha{}_\beta\nu(U,n)^\beta ]
  \nonumber\\
         &=&  \gamma(U,n) \tilde{\hat\nu}(U,n)^\alpha  
                  d \tilde p(U,n) / d\tau_{(U,n)} 
  \nonumber\\
        & &\ +  \gamma(U,n)^2 [ \nu(U,n)^2 
         \tilde D_{\rm(lie\flat)}(U,n) \tilde{\hat\nu}(U,n)^\alpha
                                            / d \tilde\ell_{(U,n)} 
  \nonumber\\
         & &\   + \gamma(U,n)^{-2}\nabla(n)^\alpha \ln N ]
\end{eqnarray} 
and in the slicing point of view
\begin{eqnarray}
  \gamma(U,n) A(U,n)_\alpha
         &=&  \gamma(U,n)  \tilde{\hat\nu}(U,n)^\alpha 
                  d \tilde p(U,n) / d\tau_{(U,n)} 
  \nonumber\\
        & &\  +  \gamma(U,n)^2 [ \nu(U,n)^2 
         \tilde D_{\rm(lie)}(U,n,e_0) \tilde{\hat\nu}(U,n)^\alpha
                                            / d \tilde\ell_{(U,n)} 
  \nonumber\\
         & &\   + \gamma(U,n)^{-2}\nabla(n)^\alpha \ln N
  \nonumber\\
        & &\ - H_{\rm(lie\flat)}(n,e_0)^\alpha{}_\beta
                  \nu(U,n)^\beta ] \ .
\end{eqnarray}

In each point of view the key difference between the original and the
conformally rescaled versions of these equations is the removal of the
gamma squared factor which multiplies the lapse derivative term, which
comes about from the difference term between the two covariant
derivatives and the identity (\ref{eq:emc2}). For purely transverse
relative accelerated motion in which the spatial acceleration lies in
the common rest subspace $LRS_U \cap LRS_u$ (as in circular motion),
then gamma times the spatial acceleration is the 4-acceleration
itself, which can therefore be expressed as the sum of the logarithmic
gradient of the lapse plus gamma squared times an ``optical
centripetal acceleration" 
\begin{equation} 
   \tilde a^{(\bot)}_{\rm(tem)}(U,u)_\alpha 
     = \tilde\nu(U,u)^2 
         \tilde D_{\rm(tem)}(U,u) \tilde{\hat\nu}(U,u)_\alpha 
                                          / d \tilde\ell_{(U,u)} 
\end{equation}
plus additional terms due to the gravitomagnetic vector force and
possible temporal derivatives. This is in some sense the decomposition
of Abramowicz, which focuses on the optical centripetal acceleration,
used to define an ``optical centrifugal force" as gamma squared times
the sign reversal of the optical centripetal acceleration \cite{a90b} 
\begin{equation}
     \tilde f^{(\bot)}_{\rm(tem)}(U,u)_\alpha
   =    - \gamma(U,u)^2 \tilde a^{(\bot)}_{\rm(tem)}(U,u)_\alpha  \ .
\end{equation}
The square of the gamma factor enters from the change of proper time
between the observer and test particle in the second derivative
defining the acceleration. 

One can also introduce the optical relative curvatures
$    \tilde\kappa_{\rm(tem)}(U,u) $
and 
\hfil\break\typeout{hfil break inserted}
$    \tilde\kappa_{\rm(lie)}(U,n,e_0) $
as the magnitude of the optical derivatives
$ \tilde D_{\rm(tem)}(U,u) \tilde{\hat\nu}(U,u)^\alpha / 
  d  \tilde \ell_{(U,u)} $
and 
$ \tilde D_{\rm(lie)}(U,n,e_0) \tilde{\hat\nu}(U,n)^\alpha / 
  d  \tilde \ell_{(U,n)} $\typeout{tildes over D omitted in original}
respectively, but all of these will only be relevant in the static
case. This leads to the ``optically straight" world lines which play a
key role in the ``reversal of the centrifugal force" discussion of
Abramowicz et al. 

Note finally that presence of the inverse square of the gamma factor
multiplying the logarithmic gradient of the lapse in the conformally
rescaled force equations above shows clearly that the
ultrarelativistic limit for geodesic motion approaches the free photon
behavior in stationary spacetimes for which the Lie derivative term is
zero. As the relative speed approaches 1 and the gamma factor becomes
increasingly large, this term drops out and the observer arclength
parametrization approaches the proper time parametrization. One is
then left with a balance of the optical centripetal acceleration and
the gravitomagnetic force as occurs for photons, as noted by
Abramowicz et al \cite{anw95} for the case of circular orbits in
stationary axially symmetric spacetimes. 

In a long review \cite{a93}, both the threading and hypersurface
points of view were mentioned in the context of stationary
axisymmetric spacetimes, and both points of view were applied to the
Kerr spacetime in separate articles \cite{pc90,ip93}, although the
ambiguity of the individual force terms in this approach was not
addressed until the gauge-fixing discussion of a later version of the
force decomposition \cite{anw93}. Although it is not easy to decipher
(complicated by sign inconsistencies and minor errors), the
subsequent presentation of the Abramowicz et al definition of
noninertial forces for general spacetimes \cite{anw95} is just the
hypersurface point of view decomposition (\ref{eq:abram}) of the
sign-reversed spatial projection of the test particle 4-acceleration
expressed in terms of the Lie total spatial covariant derivative and
the contravariant relative velocity, with the gravitoelectric force
and longitudinal and transverse relative accelerations shuffled among
themselves by the conformal transformation. They refer to the latter
two as the Euler and centrifugal forces when sign-reversed and
conformally shuffled. The gravitomagnetic tensor force is entirely due
to the expansion tensor, which they refer to as the Coriolis force
with its extra gamma factor, while their gravitational force is just
the gravitoelectric vector field itself without the gamma factor which
occurs in the gravitoelectric force. The ``ACL gauge" they have chosen
for the extension of $U^\alpha$ off its worldline in the hypersurface
point of view in order to make their derivative expressions meaningful
is just $\nabla_{\rm(lie)}(n) \hat\nu(U,n)_\alpha = 0$ (inconsistent
with their covariant derivative formula for this derivative). One
could also repeat their discussion with the equally valid but distinct
corresponding contravariant gauge condition $\nabla_{\rm(lie)}(n)
\hat\nu(U,n)^\alpha = 0$, as well as switch to the threading point of
view as alluded to in an earlier article \cite{a93}. 

None of their forces coincide with the relative forces measured by the
observers, and in the case of more general motion in which the various
forces (apart from the longitudinal relative acceleration) are no
longer all transverse, they do not lie in the local rest space of the
test world line, so their interpretation in terms of forces that would
be seen with respect to axes which comove with the test world line
(apart from proper time adjustments) are no longer valid. This
inconsistency appears in the claim that the various forces they
introduce lie in the ``comoving frame of the particle" \cite{anw93}
when in fact they lie in the local rest space of the observer (the
exception being the intersection of the two subspaces orthogonal to
the relative 2-plane of the motion). 

Perhaps one can best characterize the limitations of the approach and
its development by saying that it was born in a very special context
and then attempts were made to generalize it, rather than realizing it
as a specialization of an already general approach for arbitrary
spacetimes, the tools of which have been around for a long time, but
simply lacked a unifying umbrella. In particular, the key acceleration
potential equation $u^\beta \nabla_\beta u_\alpha = \pm \nabla_\beta
\Phi$ underlying this approach is inappropriate for nonstationary
spacetimes where an additional spatial projection is needed for the
gradient term and an additional Lie derivative of a vector potential
is needed for a threading point of view. However, this said, the
original application to static spacetimes where the optical
centripetal acceleration does reverse when the relative signed optical
curvature expression $\tilde\kappa(\phi,u)^{\hat\rho}$ associated with
the $\phi$ coordinate circles changes sign is a very beautiful
geometrization of the relative motion of massive and massless test
particles, for which credit is clearly due for its recognition and
description. 

\section{Concluding Remarks}

By implementing relatively straightforward ideas about special
relativistic space-plus-time splitting in the context of general
relativity, using a notation which allows one to examine all the
possibilities for generalizing concepts which do not have unambiguous
extensions into the more general arena, a foundation has been built
which enables one to analyse any particular problem that involves the
description of idealized observations in a given spacetime. Not only
are such questions fascinating, but their answers are often
sufficiently subtle that much confusion has arisen even in the case of
relatively simple spacetimes. Indeed the case of ``rigid rotation" in
flat spacetime itself still leads people astray in their attempts to
come to terms with such questions. 

Armed with the present tools, one can examine the traditional test
cases for investigating these ideas \cite{rinper}, namely Minkowski
spacetime in rotating coordinates, the black hole spacetimes of Kerr
and Schwarzschild, and the spacetime which was the first to
dramatically challenge our intuition about rotation in general
relativity nearly half a century ago, the G\"odel spacetime. This will
be done in a companion article, leading to a much clearer
understanding not only of these spacetimes but of the tools themselves
for studying other spacetimes. 

\section*{Acknowledgements}

We thank Remo Ruffini of the International Center for Relativistic
Astrophysics at the University of Rome and Francis Everitt of the
Gravity Probe B Relativity Mission group at Stanford University for
their support and encouragement of this work, as well as the many
relativists before us who have laid the groundwork for our analysis.
 
\appendix

\section{Conformal transformations}

\subsection{General considerations}

Following the abstract index notation of Wald \cite{wal}, suppose one
has a pair of conformally related metrics 
\begin{equation}
   \tilde g_{ab} = \sigma^2 g_{ab} \ ,
\end{equation}
each with its associated symmetric connection $\nabla$ and
$\tilde\nabla$. Then under the additional conformal transformation of
a vector 
\begin{equation}
     \tilde v^a = \sigma^{-1} v^a \ , \quad 
     \tilde v_a = \sigma v_a \ , \quad
     \tilde v{}^a \tilde v{}_a = v^a v_a  
\end{equation}
which preserves its magnitude, Eq.~(D.5) of Wald
\begin{equation}
   v^a \tilde\nabla_a v^b 
     = v^a \nabla_a v^b 
       + (2v^b v^c - g^{bc} v^dv_d) \nabla_c(\ln \sigma) \ ,
\end{equation}
is easily transformed to
\begin{eqnarray}
   \sigma^2 \, \tilde v{}^a \tilde\nabla_a \tilde v{}^b 
     &=& v^a \nabla_a v^b 
         + (v^b v^c - g^{bc} v^dv_d) \nabla_c(\ln \sigma) 
             \nonumber\\
     &=& v^a \nabla_a v^b 
         - v^d v_d P(v)^{bc} \nabla_c(\ln \sigma) \ ,
\end{eqnarray}
where the last equality is only valid in the case $v^d v_d \neq0$.
$P(v)^a{}_b$ is just the tensor which projects orthogonal to $v^a$ in
that case, necessary since the either covariant derivative of a unit
vector must be orthogonal to its direction, given that the metric is
covariant constant. Using the latter fact, one obtains the covariant
form of this equation 
\begin{eqnarray}
   \tilde v{}^a \tilde\nabla_a \tilde v{}_b 
    &=& v^a \nabla_a v_b 
           + (v_b v^a - \delta^a{}_{b} v^cv_c) \nabla_a(\ln \sigma) 
         \nonumber\\
    &=& v^a \nabla_a v_b 
           - v^c v_c P(v)^a{}_{b} \nabla_a(\ln \sigma) \ .
\end{eqnarray}
 
\subsection{Conformal transformations of spatial quantities}

For a given family of test observers with 4-velocity $u^\alpha$ and a
given timelike world line of a test particle with 4-velocity
$U^\alpha$, consider a conformal transformation of the spatial metric 
\begin{equation}\label{eq:ct}
      \tilde P(u)_{\alpha\beta} = \sigma^2 P(u)_{\alpha\beta} \ .
\end{equation}
One can introduce a new spatial covariant derivative associated with
the new spatial metric and use it to re-express the various total
covariant derivatives of the relative velocity of a test particle. For
our present purposes this is really only useful in the context of a
nonlinear reference frame where the spacetime  metric may be
conformally rescaled by the inverse square of the lapse function,
corresponding to a conformal factor $\sigma$  equal to the reciprocal
of the lapse function, $ \sigma = M^{-1}$ or $\sigma =N^{-1}$
respectively (the optical gauge), leading to the optical spatial
metric whose components in an observer-adapted frame have been denoted
by $\gamma_{ab}$ and $g_{ab}$ respectively. However, the
``antioptical" gauge $ \sigma = M$ or $\sigma = N$ respectively is
important in certain analyses of the gravitational field equations
themselves in the case of stationary spacetimes and in post-Newtonian
approximations. 

From their definitions, the spatial arclength parameter, the relative
speed, and the relative velocity unit vector must transform in the
following way 
\begin{eqnarray}
  & &      d \tilde\ell_{(U,u)} / d \ell_{(U,u)} = \sigma \ , 
              \nonumber\\
  & &      \tilde\nu(U,u) = d \tilde\ell(U,u) / d \tau_{(U,u)}
          = \sigma d \ell(U,u) / d \tau_{(U,u)}
          = \sigma \nu(U,u) \ ,           
               \nonumber\\
  & &    \tilde{\hat\nu}(U,u)^\alpha 
          = \sigma^{-1} \hat\nu(U,u){}^\alpha \ ,
\end{eqnarray}
implying that the relative velocity vector 
\begin{equation}
    \tilde\nu(U,u)^\alpha 
     = \nu(U,u)^\alpha 
     = P(u)^\alpha{}_\beta d x^\beta / d \tau_{(U,u)}
\end{equation}
is invariant.

One may also introduce the conformally rescaled energy and momentum
(per unit mass) 
\begin{eqnarray}
     \tilde E(U,u) 
           &=& \sigma^{-1} E(U,u) = \sigma^{-1} \gamma(U,u)\ ,
                  \nonumber\\
      \tilde p(U,u) &=& \sigma^{-1} p(U,u) = \sigma^{-2} 
                \gamma(U,u) \tilde\nu(U,u) \ ,
                  \\
      \tilde p(U,u)^\alpha &=& \sigma^{-2} p(U,u)^\alpha
                   \nonumber\\
      \tilde p(U,u)_\alpha &=& p(U,u)_\alpha
                   \nonumber
\end{eqnarray}
satisfying
\begin{equation}
       \tilde E(U,u)^2 - \tilde p(U,u)^2 = \sigma^{-2} \ .
\end{equation}
This choice for the conformal scaling of these last two quantities is
made so that in the special case of a Killing observer where
$u^\alpha= M^{-1}\xi^\alpha$, with $\sigma^{-1}$ taken to be the
magnitude $M$ of the Killing vector $\xi^\alpha$, then the conformally
rescaled spatial metric is the optical spatial metric and the
conformally rescaled energy is the conserved quantity $\tilde E(U,u) =
- p(U)^\alpha \xi_\alpha $ which is constant along the test particle
world line, while a Killing component of the covariant spatial
momentum remains a conserved quantity if it is not rescaled. 

Finally, for a null geodesic where the equations of motion are
conformally invariant, the new affine parameter can be defined by $d
\tilde\lambda_P / d \lambda_P = \sigma^2$ as in Eq. (D.6) of Wald,
leading to this same choice of conformal transformation for the
momentum 4-vector $P^\alpha = d x^\alpha / d \lambda_P$ and its
corresponding 1-form as  for the spatial momentum and its 1-form in
the case of a massive particle. This is easily seen from the action
which gives the equations for affinely parametrized null geodesics as
its Lagrangian equations 
\begin{equation}
      \int 
    g_{\alpha\beta}  (d x^\alpha / d \lambda_P)  
           (d x^\beta / d \lambda_P) \, d \lambda_P \ .
\end{equation}
This choice of transformation for $\lambda_P$ leaves the action
invariant. 

One may introduce a new spatial covariant derivative
$\tilde\nabla(u)_\alpha$ associated with the conformally rescaled
spatial metric by using the appropriate difference connection as in
Eqs.~(D.3) and the sign reversal of (3.1.7) of Wald \cite{wal} 
\begin{eqnarray}\label{eq:tildedelX}
 \tilde\nabla(u)_\alpha X_\beta 
   &=& \nabla(u)_\alpha X_\beta 
      - X_\gamma {\cal C}^\gamma{}_{\alpha\beta}\ ,
               \nonumber\\
 {\cal C}^\gamma{}_{\alpha\beta} 
   &=&  [2 \delta^\gamma{}_{(\alpha} \nabla(u)_{\beta)} 
           - g_{\alpha\beta} \nabla(u)^\gamma ] (\ln \sigma) \ .
\end{eqnarray}
This in turn may be used to introduce a conformally rescaled total
spatial covariant derivative of each type, by replacing the spatial
covariant derivative in the definition valid for vector fields by the
conformally rescaled derivative. For example, for a spatial covector
one has 
\begin{equation}
    \tilde D_{(\rm tem)}(U,u) X_\beta / d \tau_{(U,u)} 
          = D_{(\rm tem)}(U,u) X_\beta / d \tau_{(U,u)} 
           - X_\gamma {\cal C}^\gamma{}_{\alpha\beta} \nu(U,u)^\alpha \ .
\end{equation}

One can easily re-express the three-acceleration or rate of change of
spatial momentum in terms of this new derivative, using an immediate
consequence of Eq. (\ref{eq:tildedelX}) 
\begin{equation}
 X^\alpha \tilde\nabla(u)_\alpha X_\beta 
  = X^\alpha \nabla(u)_\alpha X_\beta 
         - X^\alpha X_\alpha \nabla(u)_\beta( \ln\sigma) \ .
\end{equation}
For example, for a congruence of test particle world lines one would
have 
\begin{eqnarray} 
  D_{(\rm tem)}(U,u) p(U,u)_\beta / d \tau_U
     &=& E(U,u) \nabla_{(\rm tem)}(u) p(U,u)_\beta
         + p(U,u)^\alpha \nabla(u)_\alpha p(U,u)_\beta \nonumber\\
     &=&   \tilde D_{(\rm tem)}(U,u) \tilde p(U,u)_\beta / d \tau_U
         + p(U,u)^2 \nabla(u)_\beta (\ln\sigma) \ ,
       \nonumber\\ \hbox{} 
\end{eqnarray}
where
\begin{eqnarray} \label{eq:BDp}
    \tilde D_{(\rm tem)}(U,u) \tilde p(U,u)_\beta / d \tau_U
     &=& E(U,u) \nabla_{(\rm tem)}(u) p(U,u)_\beta
           + p(U,u)^\alpha \tilde\nabla(u)_\alpha p(U,u)_\beta 
          \nonumber\\
     &=&   D_{(\rm tem)}(U,u) p(U,u)_\beta / d \tau_U
              - p(U,u)^2 \nabla(u)_\beta (\ln\sigma)
       \nonumber\\ \hbox{} 
\end{eqnarray}
defines the rescaled total spatial covariant derivatives for either a
congruence of test particle world lines or a single such world line
respectively. Raising the index on these equations introduces an extra
term from the temporal derivative of the conformal factor. These
formulas yield the results for the threading and hypersurface points
of view in the optical gauge for $u=m,n$, while replacing $n$ by
$n,e_0$ in the appropriate places in the hypersurface Lie form of
these equations yields the slicing version. 

One can similarly transform the derivative of the unit velocity vector
needed to evaluate the relative centripetal acceleration, leading to
the conformally rescaled quantities corresponding to the relative
curvature and radius of curvature and the relative centripetal
acceleration. If $\hat\nu(U,u)^\alpha$ were actually a vector field on
spacetime rather than being defined only along a single world line,
one could decompose its spatially projected intrinsic derivative in
the following way 
\begin{eqnarray}
  &&  D_{\rm(tem)}(U,u) \hat \nu(U,u)_\beta / d \ell_{(U,u)} 
 \nonumber\\
  &&\quad = [ 1 / \nu(U,u) ] 
           D_{\rm(tem)}(U,u) \hat \nu(U,u)_\beta / d \tau_{(U,u)} 
 \nonumber\\
  &&\quad = [ 1 / \nu(U,u) ] 
        [ \nabla_{\rm(tem)}(u) 
                + \nu(U,u) \hat \nu(U,u)^\alpha \nabla(u)_\alpha ]
                \hat \nu(U,u)_\beta \ .
\end{eqnarray}
Now re-express this in terms of the conformally rescaled quantities
using the results of the first section of the appendix for the unit
velocity vector to re-express the spatial covariant derivative in it
in terms of a conformally rescaled derivative
$\tilde\nabla(u)_\alpha$. One finds 
\begin{eqnarray}
  &&  D_{\rm(tem)}(U,u) \hat \nu(U,u)_\beta / d \ell_{(U,u)} 
           \nonumber\\
  &&\quad = \tilde D_{\rm(tem)}(U,u) \tilde{\hat\nu}(U,u)_\beta 
                                         / d \tilde\ell_{(U,u)} 
           \nonumber\\
  &&\quad\  + P_u(U,u)^{(\bot)}{}_\beta{}^\alpha 
                          \nabla(u)_\alpha \ln\sigma
              - \hat\nu(U,u)_\beta / \nu(U,u) \,
                           \nabla_{\rm (fw)}(u) \ln\sigma \ ,
\end{eqnarray}
where the first of the following equalities
\begin{eqnarray}
  &&  \tilde D_{\rm(tem)}(U,u) \tilde{\hat\nu}(U,u)_\beta  
                                     / d \tilde\ell_{(U,u)} 
          \nonumber\\
  &&\quad = [ 1 / \tilde\nu(U,u) ] 
       [ \nabla_{\rm(tem)}(u) 
         + \tilde\nu(U,u) \tilde{\hat\nu}(U,u)^\alpha 
                   \tilde\nabla(u)_\alpha ] 
                       \tilde{\hat\nu}(U,u)_\beta 
  \nonumber\\
  &&\quad  = D_{\rm(tem)}(U,u) \hat \nu(U,u)_\beta / d \ell_{(U,u)} 
  \nonumber\\
  &&\quad\ - P_u(U,u)^{(\bot)}{}^\alpha{}_\beta 
                              \nabla(u)_\alpha \ln\sigma
              + \hat\nu(U,u)_\beta / \nu(U,u) \,
                              \nabla_{\rm (fw)}(u) \ln\sigma 
\end{eqnarray}
defines the equivalent action of the new derivative on a congruence of
test particle world lines, while the second defines it for a single
such world line. 

The magnitude of this conformal derivative of the conformal unit
velocity defines the conformally rescaled relative curvature and its
reciprocal the conformally rescaled radius of curvature, for the
ordinary and corotating Fermi-Walker cases. Multiplying the last
relationship by the conformal square of the velocity gives the
conformally rescaled relative centripetal acceleration and its
relationship to the original one as long as the total spatial
covariant derivative is orthogonal to the relative direction of motion
\begin{eqnarray}
  \tilde a_{(\rm tem)}^{(\bot)}(U,u)_\beta
   &=& \sigma^{2} [ a_{(\rm tem)}^{(\bot)}(U,u)_\beta
         -\nu(U,u)^2  P_u(U,u)^{(\bot)}{}^\alpha{}_\beta
                     \nabla(u)_\alpha \ln\sigma
        \nonumber\\
   &&\quad\ + \nu(U,u)_\beta \nabla_{\rm (fw)}(u) \ln\sigma ] \ ,
              \quad {\rm\scriptstyle tem = fw,\, cfw} \ .
\end{eqnarray}
Note that unless the observer temporal derivative of the conformal
factor is zero, the ``conformal derivative" of the unit velocity is
not orthogonal to the unit velocity itself, even if it is before the
conformal rescaling. The same is true of the conformally rescaled
relative centripetal acceleration. 

In the stationary case with $\sigma$ chosen for the optical gauge,
this temporal derivative vanishes while the spatial derivative
produces the gravitoelectric field, and this relationship can be
rewritten in the form (using first Eq.~(\ref{eq:emc2}) and then
Eqs.~(\ref{eq:aU})--(\ref{eq:Ftem}))
\begin{eqnarray}
  &&  \gamma(U,u)^2\sigma^{-2}\tilde a_{(\rm tem)}^{(\bot)}(U,u)_\beta
        \nonumber\\
  &&\quad = P_u(U,u)^{(\bot)\, \alpha}{}_\beta 
           \{ 
  \gamma(U,u)^2 [ a_{(\rm tem)}^{(\bot)}(U,u)_\alpha - g(u)_\alpha ]
             + g(u)_\alpha \} 
\\
  &&\quad = P_u(U,u)^{(\bot) \alpha}{}_\beta [ a(U)_\alpha +  g(u)_\alpha
     + \gamma(U,u)^2 H_{\rm(tem)}(u)_{\alpha\delta} \nu(U,u)^\delta]
           \ .
             \nonumber
\end{eqnarray}
Thus if in addition the transverse gravitomagnetic force is zero, one
sees that the optical relative centripetal acceleration changes sign
when the transverse spatial projection of the test particle
acceleration just balances the transverse gravitoelectric field, i.e.,
exactly opposes the transverse observer acceleration
\begin{eqnarray}
  \gamma(U,u)^2\sigma^{-2}\tilde a_{(\rm tem)}^{(\bot)}(U,u)_\beta
   &=& P_u(U,u)^{(\bot) \alpha}{}_\beta [ a(U)_\alpha - a(u)_\alpha ]
           \ ,
\\
  \gamma(U,u)^2  a_{(\rm tem)}^{(\bot)}(U,u)_\beta
   &=& P_u(U,u)^{(\bot) \alpha}{}_\beta [ a(U)_\alpha 
                                      - \gamma(U,u)^2 a(u)_\alpha ]
           \ ,\nonumber
\end{eqnarray}
where for comparison the analogous relation for the relative
centripetal acceleration itself is given (just the transverse analog
of the static case longitudinal acceleration relation (3.13) of
\cite{rok}). For static circular orbits, the optical relative
centripetal acceleration then becomes outward pointing when the
transverse test particle 4-acceleration becomes larger in magnitude
than the transverse test observer 4-acceleration. This is the famous
effect of the ``reversal of the centrifugal force" motivating the work
of Abramowicz et al. The additional squared gamma factor in the
relative centripetal acceleration compared to the optical relative
centripetal acceleration prevents the reversal in the former case. 




\section*{Corrections}

This reformatted version contains two reference publication updates and one typo correction:
line 4 after 15.10 (top of p.~31 in the original article) where
``the optical derivatives $\tilde D \ldots$" should have the two uppercase D derivative symbols with an over tilde.

\end{document}